# A Hierarchical Energy-Based Model for Multimodal Cognition

**Subir Varma**

*General Cognitics, subir@gencog.com*

August 2026

## Abstract

We propose IM-LEPP (Integrated Multimodal Latent Energy-based Predictive Processing), a hierarchical, energy-based model of multimodal cognition that extends a previously proposed single-modality perceptual model (LEPP) to integrate vision and language. Building on the view that generative neural networks can serve as effective theories of cognitive dynamics, in the way statistical mechanics relates to thermodynamics, IM-LEPP models cognition as the flow of latent states through learned energy landscapes rather than as an explicit account of neural circuitry. The architecture is a hub-and-spoke hierarchy grounded in the controlled semantic cognition framework of Lambon Ralph et al., in which predictive-coding pipelines for individual visual objects and scenes, and for discrete linguistic units (phonemes/characters, then words), converge on a shared amodal hub modeled on the anterior temporal lobe (ATL). Each pipeline's diffusion-based prediction is conditioned by, rather than overwritten by, the current hub state, preserving pipeline-specific identity while letting every prediction reflect the full multimodal context. We show that this architecture gives a mechanistic account of attentional phenomena such as inattentional blindness and Necker-cube bistability, and that its mathematical structure recovers or motivates independently established findings in psycholinguistics, including surprisal theory, the N400/P600 ERP components, and garden-path reanalysis, alongside a falsifiable point of contrast with transformer-based language models on trajectory-sensitivity in next-word prediction. We discuss implications for the data efficiency of human language acquisition relative to large language models, outline a semantic/episodic memory subsystem, situate the model against related frameworks including predictive coding, the free-energy principle, joint-embedding predictive architectures (JEPA), and Hierarchical Temporal Memory, and propose a set of concrete experimental predictions to test the model's central claims.

## 1 Introduction

This paper proposes an hierarchical computational model for semantic cognition in the brain. The model, that we call Integrated Multimodel Latent Energy based Predictive Processing or IM-LEPP builds on the previous paper that proposed the LEPP model for perception as a process of flow on energy landscapes. The IM-LEPP model extends this idea and builds a model for cognition that integrates both perception and language processing. The IM-LEPP architecture allows for additional sensory modalities to be integrated into the model in a straightforward manner.

The IM-LEPP design is based the following core idea that was introduced in the paper Generative AI as an Effective Theory of Cognition: Modern generative neural networks should be understood not as mechanistic models of neural implementation, but as effective theories of cognitive dynamics operating at the level of learned energy landscapes. In this view, cognition is understood as the temporal evolution of perceptual and cognitive states through a learned energy landscape, rather than as the direct consequence of an explicitly modeled neural circuitry. Just as statistical mechanics explains why thermodynamics provides an effective description of macroscopic matter without explicitly modeling every molecular interaction, modern generative AI may provide effective descriptions of cognitive dynamics without modeling the underlying neural circuitry. Leveraging this insight, IM-LEPP models the brain at the level of energy landscapes rather than at the circuit level.

The IM-LEPP architecture is also based on the Inference-Prediction-Generation framework for cognitive processing that was used in previous paper, which is referred to as predictive processing. Both vision and language have their own inference, prediction and generation pipelines that operate using the predictive processing model. The inference pipeline results in latent states that reflect the latest vision or language data. This latent state is sent into a central hub, where it gets integrated with latent states of other sensory modules and results in an integrated state. The integrated state is then sent back to the original sensory modality and fed into a prediction module that then predicts the next latent state candidate. The prediction is then used to generate the next percept (in the case of vision) or word (in the case of language). It also gets modified into a new latent state as new sensory data comes in, and the cycle repeats.

This framework allows the vision and language processes to evolve independently of each other, and operate at different time scales. This aspect is important since vision sensory data is received almost continuously, while language data in the form of words is separated by a few hundred milliseconds. The central integrated state gets invoked asynchronously as new data comes in from either module. This design allows for the various modalities in the brain to influence one another as their states evolve. The IM-LEPP model also allows for continual learning, since all information required to update the parameters of the various modules is available locally at each step of the inference, generation and prediction pipeline. All the modules proposed as part of IM-LEPP framework operate using the principle of energy minimization, and all communications between states is local in nature. Thus it serves as a plausible model for the brain, at the algorithmic level (i.e., Marr's level 2). As pointed out in the previous paper, it is likely that the brain also operates using the principle of energy minimization, but at the level of its connectome, which is invisible to us.

This paper proposes a solution for the binding problem in the brain, i.e., the problem of integrating various modalities together into a common representation. It proposes that binding is accomplished through the use of predictive coding pipelines for individual modalities. This is in agreement with what is known about how binding between different sensory modalities happens in the anterior temporal lobe (ATL) of the brain, as described in the paper by Ralph et al. 2017.

We introduce an hierarchical hub and spoke model for vision, in which the base level is made up of predictive processing pipelines for objects that are in the field of vision as well for the scene as a whole. The individual latent states from these modules gets integrated at the vision hub into a combined vision state, which in turn gets integrated with other modalities at the central hub. Thus each object has its own individual model in this framework, and evolves in time asynchronously to the other objects. However the predictions from each object module reflect the presence of other objects as well as other modalities such as language and the mechanism for this is described in this paper.

We also introduce a predictive processing model for language which also has a two level hierarchical structure. The lower level consists of a predictive processing pipeline in which the sensory data consists of discrete phonemes (for sound) or discrete characters (for reading) coming in through the auditory or visual systems. This in turn is used to drive a higher level predictive processing pipeline that operates at the word level, and whose job is to predict the latent representation for next word. However before the next word prediction takes place, the word latent is sent to the central hub that integrates it with the vision latent state (and perhaps other modalities). Subsequently is gets fed back to prediction module of the word pipeline, and the resulting state is used to generate the latent for the next word. Finally this is fed back into the phoneme level pipeline to generate the next sound. Thus language learning and generation in IM-LEPP is based not just on word level co-occurrence statistics, but is a function of everything else that is happening in other sensory modalities which has implications for the speed of language acquisition in children.

There are several points of distinction when comparing the IM-LEPP model with modern LLMs, which also seem to be carrying out similar cognitive functions:

- There is a fundamental difference in learning between IM-LEPP model and LLMs. LLMs learn by consuming a huge amount of data in the training stage and then use the resulting frozen model for all their inference operations, which is very different from the way the IM-LEPP model learns. IM-LEPP gets trained on a continuous basis as new sensory data comes in, so that continual learning co-exists with inference and prediction, and this is also the way that brains operate.
- The IM-LEPP model introduces an integrated system state that includes language and vision and can be extended with more modalities. Thus the language generated by the model is grounded by both visual and linguistic data. It is thought that LLMs also maintain an internal latent state that is not visible externally, however this state is entirely a function of language statistics and is not grounded in real world sensory data and this results in the well known symbol grounding problem, i.e., LLMs see only the linguistic symbols, never what those symbols are about. The integration of vision and language into a shared common state in IM-LEPP constitutes a solution to this problem.
- IM-LEPP gives us direct access to the latent states through which all generation is done which can be used to influence the models output. Something equivalent can be done with LLMs by probing the transformer's middle layers. In IM-LEPP the latent states used for prediction and generation is made explicit by the model while it remains hidden in LLMs.

This paper makes the following contributions:

- We propose a hub-and-spoke integration architecture for multiple sensory modalities, with modality-specific predictive processing pipelines converging on a shared latent state via predictive coding, solving both the binding problem and asynchronous multi-rate updating across modalities operating on different timescales.

- We propose an hierarchical hub and spoke model for vision, in which a central hub integrates latent states from individual objects in the field of vision as well as for the scene as a whole. Each object has its own predictive processing pipeline and its latent state evolves as a function of its own object based model. The model includes mechanisms for attentional suppression (through precision reduction), saliency-gated instantiation of new object pipelines, and open-loop prediction for objects outside current attentional focus. Together these aspects of the model give a mechanistic account of phenomena including in-attentional blindness and change blindness. We show how the energy based predictive mechanism can be used to explain the bi-stable visual states in the Necker cube.
- We propose a predictive processing model for language with parallel phoneme-based and character-based input channels, each feeding a two-level (character/phoneme at the lower level and words at the higher level) hierarchy, including an extension of predictive coding to discrete/categorical data and a proposed biologically plausible implementation of the categorical readout via divisive normalization. The next word prediction in this model is based on an integrated semantic level latent state that takes the other sensory modalities into account.
- We demonstrate that the model's own mathematical structure recovers or motivates several independently established findings in psycholinguistics and neurolinguistics, including surprisal theory, the N400 and P600 components, and garden-path reanalysis. We also discuss a specific, falsifiable point of contrast with transformer-based language models regarding trajectory-sensitivity in next-word prediction (via comparison with Barenholtz, 2026). We show how the model provides some insight into the problem of how children are able to acquire language abilities using much less training data as compared to LLMs.
- We propose a model for the memory subsystem that distinguishes semantic (object-model) and episodic memory, with associative (Hopfield-style) retrieval, a proposed novelty and valence-gated storage criterion linking hippocampal and amygdalar mechanisms, and integration into the central ATL hub.

## 2 The Hub and Spoke Model for Cognition

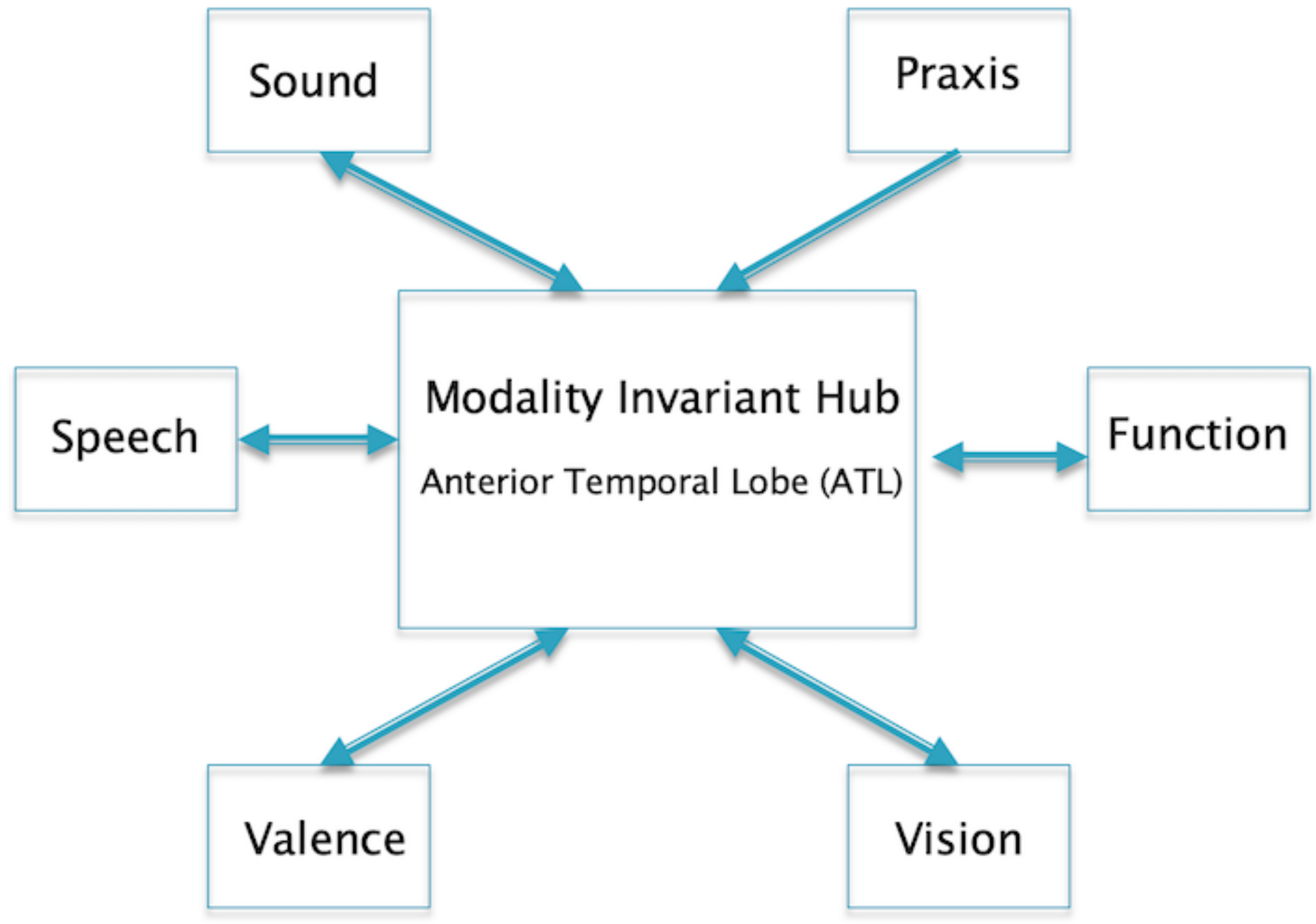

Figure 1: The hub and spoke model for cognition.

The hub and spoke model for cognition was proposed by Ralph et al. in 2017, and it was based on key findings from a decade of research into neurocognitive and neurocomputational underpinnings of the brain's semantic cognition abilities. Based on this data, they proposed the hub and spoke model of semantic representation that is shown in the above figure, which was based on the accounting of patterns of impairment that are observed in some semantic disorders. This model assimilated two important existing ideas: (a) The model assumes that multimodal verbal and non-verbal experiences provide the core ingredients for constructing concepts and these information sources are encoded in modality specific cortices distributed across the brain, (b) The model proposes that that cross modal interactions for all modality specific sources of information are mediated by a single transmodal hub that is situated bilaterally in the anterior temporal lobe (ATL) area of the brain. The hub and spoke model was suggested by the observation that individuals with semantic dementia (SD), that is characterized by atrophy centered in the ATL, show semantic impairments across all modalities.

This model solves the problem of how the information relevant to a given concept is experienced across all different verbal and sensory modalities. For example if see an image of a dog, then we are able to reproduce the sounds, names, valence (positive or negative affective value that originates in the amygdala), and other types of information that are associated with the animal. This implies that the ATL hub forms generalizable semantic representations for a dog that are shared across all modalities.

This paper develops a subset of Ralph's six modalities (namely models for vision, speech and valence) as a proof concept, with models for sound, praxis (motor/action semantics) and functional knowledge left for future work.

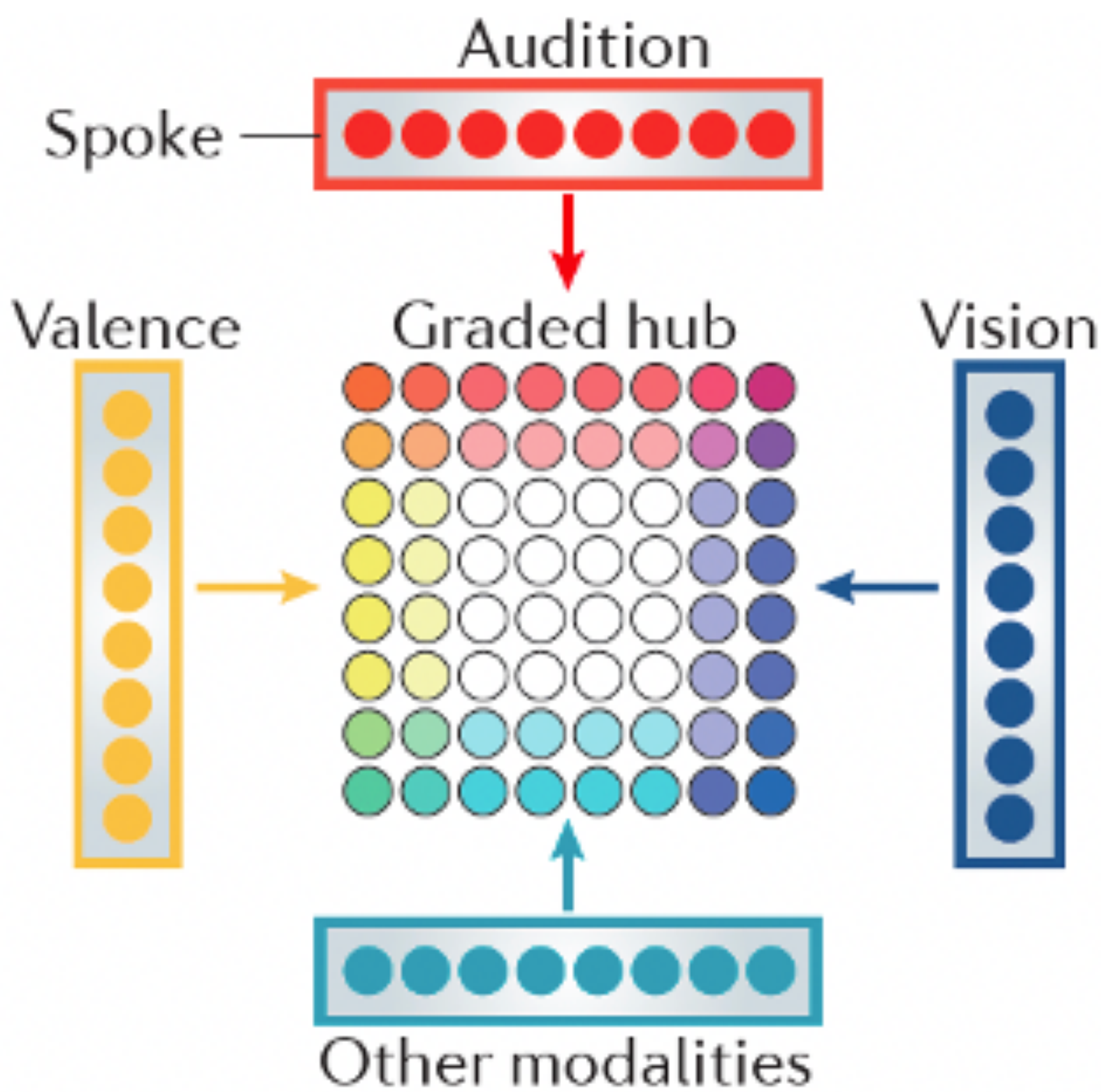

Figure 2: This figure illustrates graded pattern of semantic representation at the central ATL hub (from Ralph et al. 2017).

There is a sub-region within the ATL called the ventral-ventrolateral ATL that serves as the cross modal center point of the hub for multimodal naming and comprehension. Based on experimental data, the model proposed that semantic representation function varies in a graded manner across the the ATL subregions. The 8x8 unit grid of colored circles in the above figure represents the ATL hub with reciprocal connectivity to the modality specific spokes. The contribution of the hub units to the semantic representation is graded reflecting a varying pattern of connectivity to the spoke layers. At the center point, there is equal weighted connectivity to all inputs, thus resulting in an evenly transmodal representation, as shown by the white color. Later in this paper we present a model for the ATL hub that is based on the integration of several multi-state predictive coding pipelines that naturally reproduces this graded pattern.

It is worth noting that the status of the unified ATL hub hypothesis is more actively debated than the original 2017 formulation might suggest.
Recently, a study by Cui et al. 2025 combining voxel-based morphometry, diffusion-weighted imaging, and resting-state fMRI in 33 patients with semantic dementia found a specific functional dissociation within the proposed hub itself, namely that left ATL selectively supported abstract verbal semantic processing, while left fusiform gyrus contributed to more general semantic processing spanning verbal and non-verbal domains. Together, these findings motivate a live debate over whether cross-modal semantic integration relies on a single amodal hub located in ATL, or on a distributed network of several specialized regions, that includes the ATL together with posterior middle temporal gyrus and fusiform gyrus, with each region handling a more specific kind of integration.

This more nuanced picture is a better match to the architecture proposed in this paper than a single monolithic hub would be: the IM-LEPP model proposed here already comprises several distinct, specialized integration sites at different levels of the hierarchy (the vision hub, the language hub, and the central ATL-type hub), rather than a single region performing all integration at once. The multi-hub structure developed here can therefore be read as a computational instantiation of the distributed-hub position in this debate, rather than as a simplification of Ralph et al.'s original single-hub proposal.

## 3 The IM-LEPP Model for Vision and Language Integration

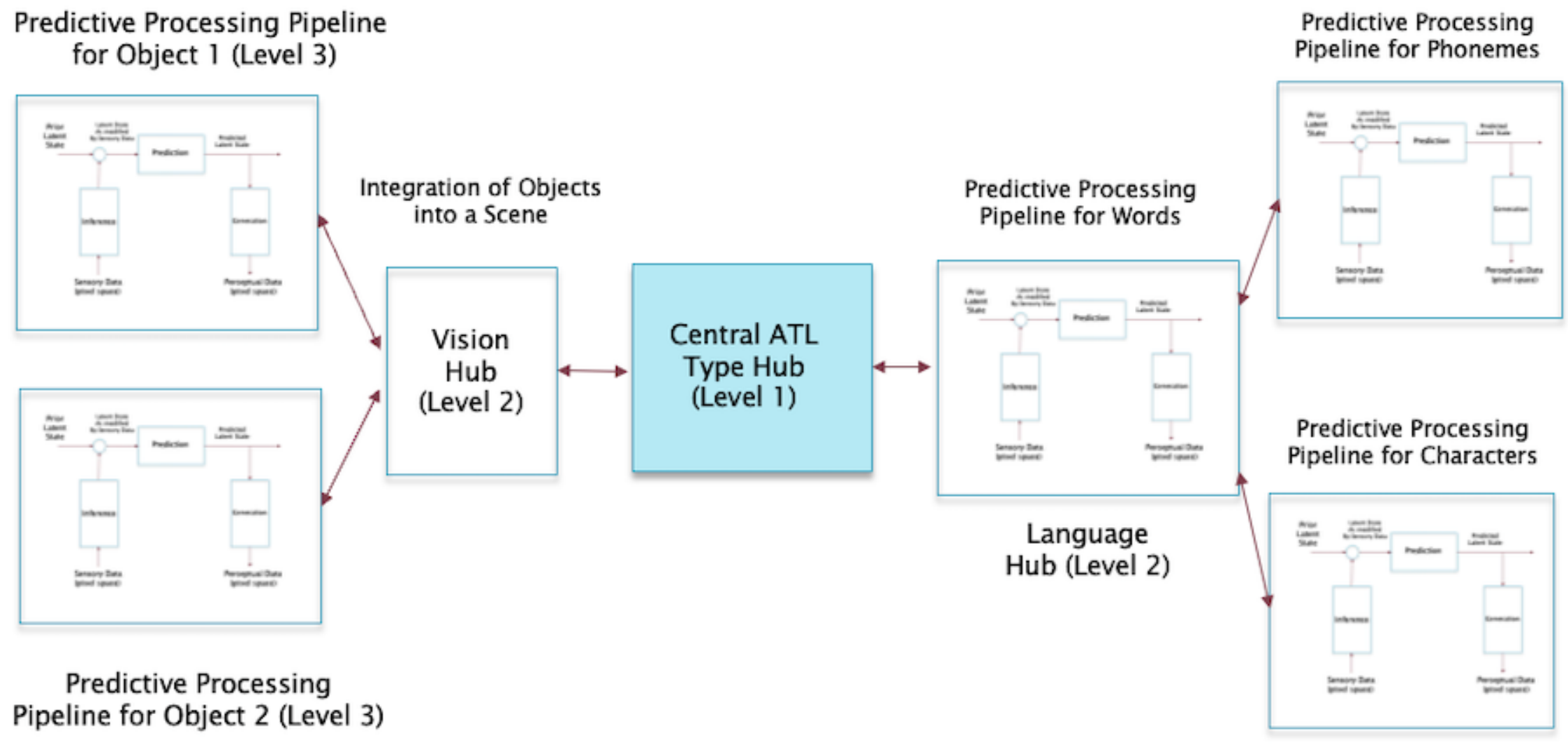

Figure 3: The IM-LEPP model

We propose a model for integration of visual and language modalities based on the hub and spoke model, which we call integrated multimodal latent energy-based predictive processing or IM-LEPP (see above figure). This model extends the LEPP model that was described in the previous paper in several ways: It proposes a hierarchical spatial integration structure for the vision model, it introduces a hierarchical temporal integration structure for language, and finally it proposes how the two may be integrated together to create a common representation. The IM-LEPP model has the following features:

- There is a central ATL type hub at level 1 that integrates representations coming in from the vision and language hubs. Note that the communication between the central hub and the vision and language hubs is bi-directional, so that not only do the spoke hubs influence the representation in the central hub, but they in turn are influenced by the information coming from the central hub.
- The vision hub itself has a two level structure. The central vision hub at level 2 integrates information coming from several simultaneously active level 3 predictive processing pipelines. There is a level 3 pipeline for each of the objects in the scene, as well an always-on pipeline for the scene itself, and all these get integrated at the level 2 vision hub. The representation of each of these pipelines evolves asynchronously in time and the level 2 hub integrates the latest information from each individual object and sends it up to the central level 1 hub. Note that the per object pipelines come and go depending on which objects are currently in the field of vision, while the scene level pipeline is always active.
- The per-object level 3 predictive processing pipelines operate according to the inference-prediction-generation framework that was used for the LEPP model. The system state that results from the inference module in this pipeline is sent to the level 2 vision hub for integration with the states of all the other objects in the scene. This combined representation in turn gets integrated with representations from other modalities in the multimodal level 1 hub. The integrated level 1 representation in turn is fed back to the prediction module in the level 3 object predictive processing pipelines, and the result is used to generate the next percept. Note that these percepts take all the other objects that are in the scene into account (as well as other modalities), by virtue of this architecture.
- The language sub-system also has a two level hierarchical structure, however the hierarchy is in time rather than in space. At the lower level of the hierarchy at level 3 are predictive processing pipelines that operate at the discrete phoneme level (in the case of spoken language) or at the character level (in the case of reading). This level incorporates an inference-prediction-generation modules whose job is to predict the next phoneme or character. Note that unlike the case for vision, only one of the level 3 pipelines is active at any one time; moreover, within whichever pipeline is active, phonemes or characters necessarily arrive one at a time in sequence rather than simultaneously, unlike the multiple objects that can be concurrently present in a visual scene. Together these account for the temporal, rather than spatial, character of the language hub. The latent representation from this level is sampled at certain discrete instants that contain representations for whole words, and these are fed as input into a word level predictive processing pipeline at level 2. The next word latent prediction done at this level is influenced by the state of the central ATL hub and thus gets modified by information from the other modalities, and ultimately gets sent to the level 3 hub to generate percepts. If the phoneme hub is active then it generates percepts in the form of sound or if the character hub is active then it generates percepts in the form of written text.

Thus the IM-LEPP model paints a picture in which there are number of distributed, predictive processing pipeline modules in the brain, that are individually responsible for predictions in the modality they are tracking. Hence the prediction operations happens in a distributed manner, while central hubs at level 1 and level 2 are responsible for integrating the lower level representations, and in turn feeding them back to the predictive processing pipelines. All state changes in this model at the various pipelines and hubs are based on the principle of the energy minimization, and thus provide a plausible model for the brain's operation at Marr's level 2 as noted in the Introduction.

## 4 A Hub and Spoke Model for Vision

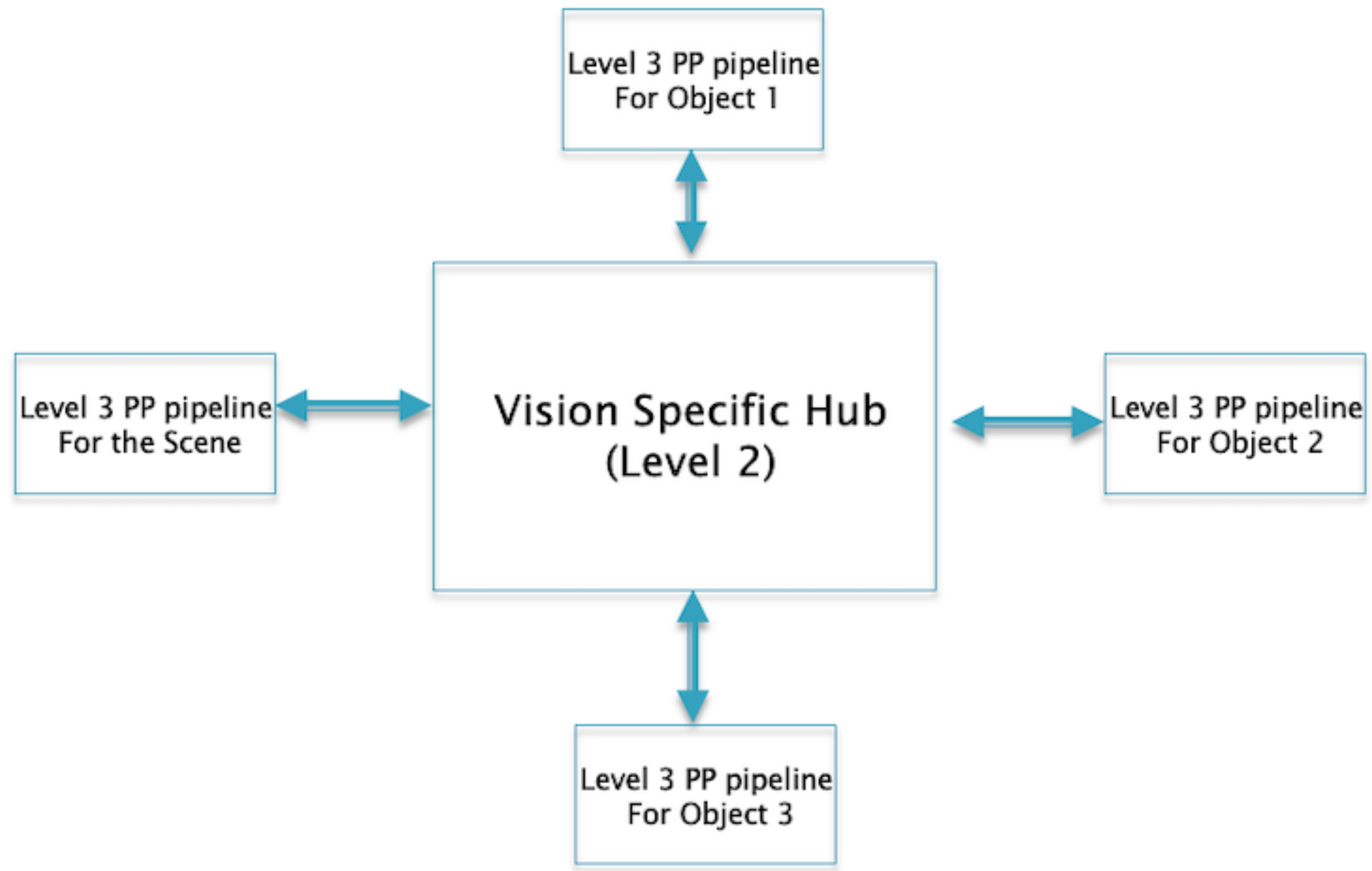

Figure 4: Hub and Spoke Model for Vision: Integration of Predictive Processing Pipelines for Multiple Objects as well as a Pipeline Representing the Scene

This section has a more detailed description of the vision model's operations at levels 2 and 3. Modern generative AI systems process images at the level of pixels, and learn prediction models for how these pixels evolves with time. The brain on the other hand is thought to decompose a scene at the object level, and then model the temporal evolution of the individual objects. For example in a road scene, there are models for pedestrians and vehicles as well as a model for the background containing the sky, foliage etc. The brain tracks each object individually, and then integrates all the representations to create the scene that we see in front of us. At the same time the brain is thought to have an always-on pipeline that tracks the representation for the scene as a whole, and thus operates more like image generative pipelines used in generative AI. The above figure shows the IM-LEPP model for vision that uses a similar hierarchical structure with both object level and whole scene pipelines included.

The IM-LEPP model proposes that the brain stores models for most of the objects that we encounter in our lives in memory, and a subset of these models is invoked and attached to the vision hub depending on the objects that are in the field of vision. If a new object is encountered, then its model is not trained from scratch, but instead it builds on an existing model that is similar to it. This allows the model to learn using much fewer training examples compared to AI systems which solely depend on using pixel level correlations. At the same time the system maintains an always-on pipeline for the scene as a whole, that is continuously active on whatever visual input is currently present, with no gating step at all. That gives the system an immediate, always-available first-pass scene representation, while individual object pipelines get instantiated and refined more gradually and selectively. This is a concrete mechanistic explanation for why scene gist is available within a fraction of a second (Potter, 1976; Thorpe et al., 1996).

Direct neural evidence exists for the individuated, per-object representations that the level 3 pipelines are intended to model. The lateral occipital complex (LOC) is well established as object-selective cortex, responding broadly to individual objects as opposed to scenes or scrambled controls, and finer-grained subdivisions within object-selective cortex show category-specific selectivity for particular object classes. For example the fusiform face area (FFA) responds selectively to faces (Kanwisher, McDermott, & Chun, 1997), while the extrastriate body area (EBA) responds selectively to human bodies and body parts (Downing, Jiang, Shuman, & Kanwisher, 2001). This neural evidence for discrete, category-specific object representations is complemented by behavioral evidence for the persistence and individuation of such representations over time. Object file theory (Kahneman, Treisman, & Gibbs, 1992) demonstrates that the visual system maintains a distinct, continuously updated representation for each individually attended object, consistent with the proposal that each tracked object is assigned its own persistent level 3 pipeline.

Evidence for the level 2 vision hub is more differentiated, and points to two distinct integrative mechanisms rather than one. A set of scene-selective regions, namely the parahippocampal place area (PPA) (Epstein & Kanwisher, 1998), the retrosplenial complex (RSC) (Maguire, 2001), and the occipital place area (OPA) (Dilks, Julian, Paunov, & Kanwisher, 2013), respond selectively to holistic scene layout, extracting global spatial and geometric structure (openness, boundary, expansion) largely independently of the number or identity of individual objects present, consistent with the fast, layout-based scene pipeline discussed above. Separately, and in a different region, MacEvoy and Epstein (2011) found that multi-voxel activity patterns evoked by whole scenes in lateral occipital cortex could be well predicted from the average of the patterns evoked by each scene's constituent objects presented in isolation. This is direct evidence that scene representations in this region are genuinely constructed by integrating individual object-level representations, rather than being computed independently of them. Together, these findings support the dual-route proposal above: a fast, layout-based route (PPA/RSC/OPA) operating largely independently of object identity, and a separate, object-integrative route (in LOC) whose representations are literally composed from individual object-level signals, corresponding respectively to the whole-scene pipeline and the object-integration mechanism already specified for the level 2 hub.

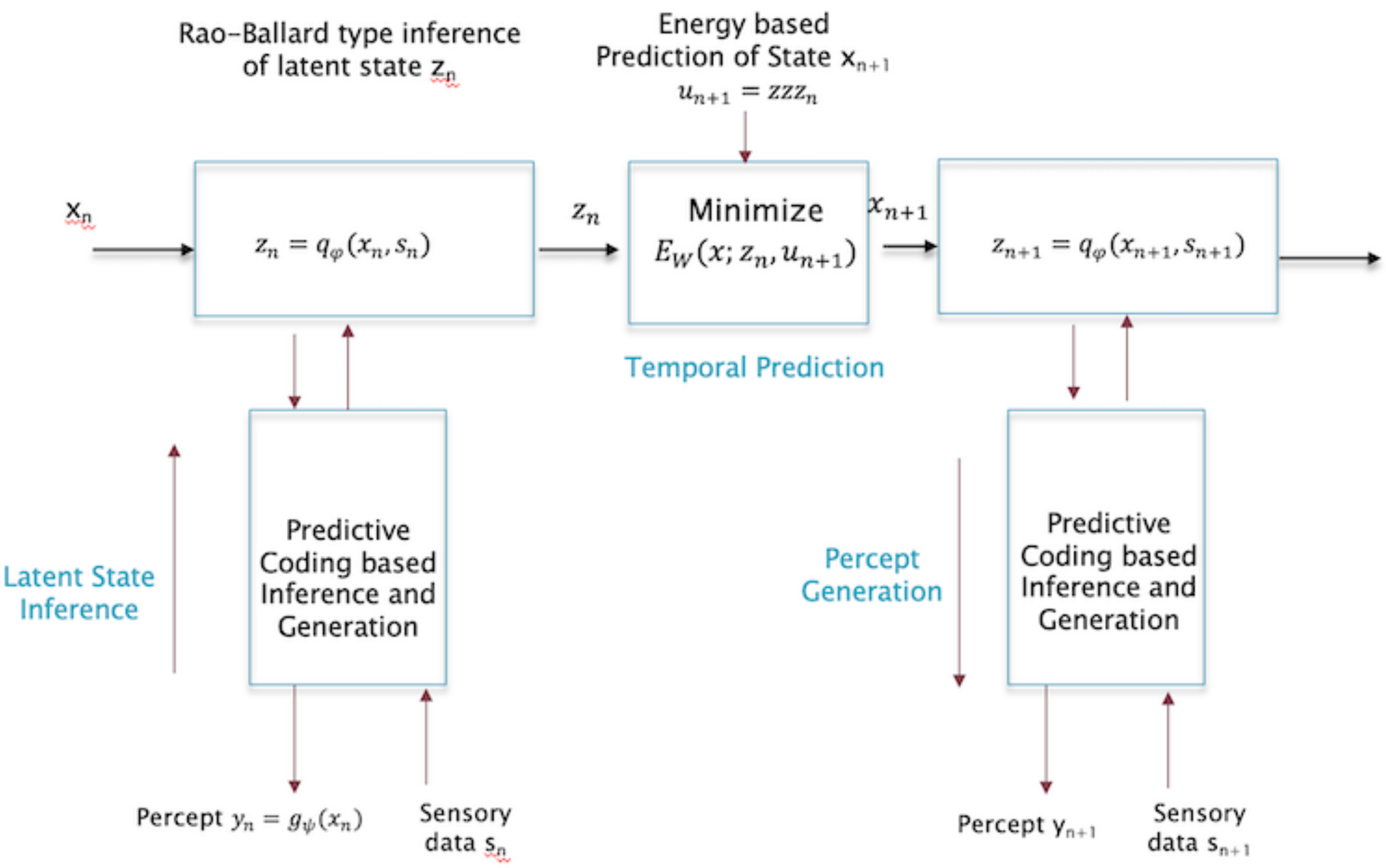

Figure 5: Predictive Processing Pipeline for an Individual Object (or Scene) at Level 3

The level 3 predictive processing pipeline for an individual object (or scene) is shown in the above figure, and follows the LEPP design, with inference, prediction and generation modules. As new sensory data $s_n$ comes in, it is integrated into the latent state $z_n$ for the object by using predictive coding as described in Varma. Inference and generation operate by using the principle of minimization of predictive coding energy $E_{PC}$ given by

$$E_{PC} = \frac{1}{2}\epsilon_y^T \Pi_y \epsilon_y + \frac{1}{2}\epsilon_z^T \Pi_z \epsilon_z$$

Thus the latent state is estimated so as to reduce the error $\epsilon_y$ between the generated value and the sensory data, plus the error $\epsilon_z$ that tracks how far latent state strays from its prior predicted value. The predictive coding based inference results in a latent state $z_n$ that is fed into the temporal prediction module and this results in a prediction $x_{n+1}$.

The temporal prediction module used throughout this paper is the LEPP diffusion-based EBM described in Varma (2026). It is modeled by using a multi-stage diffusion model that minimizes an energy $E_W(x; z_n, u_{n+1} = zzz_n)$ by stochastically annealing the state $x$ of the system using Langevin sampling, until it reaches a state $x_{n+1}$ of low energy. Hence the energy landscape reflects the latest sensory data $s_n$ through the state $z_n$, as well as the state $zzz_n$ at the shared level 1 hub. The energy landscape changes over time as new sensory data arrives and the state of the shared hub changes, thus changing the resulting minimas. This is illustrated in figure 8 in [Varma].

In order to generate the shared hub state $zzz_n$, the object level latent state $z_n$ is sent to the level 2 vision hub and this results in the integrated latent representation $zz_n$ that takes the presence of other objects as well as the overall scene into account. This in turn is sent to the multimodal level 1 hub, resulting in the latent representation $zzz_n$ which takes the other modalities into account. This state information is then fed back into initial object predictive processing pipeline where it serves as a conditioning variable to modify the energy function to $E_W(x; z_n, u_{n+1} = zzz_n)$. Note that as a result of this integration with higher level hubs, the prediction $x_{n+1}$ incorporates information about other objects in the scene as well other modalities that may be relevant such as valence. Next $x_{n+1}$ is used to generate the next percept $g_\psi(x_{n+1})$, and is also used to kick off the inference phase of the predictive processing pipeline by comparing it with the new sensory data $s_{n+1}$. This subsequently results in a new object latent state $z_{n+1}$ and the cycle repeats.

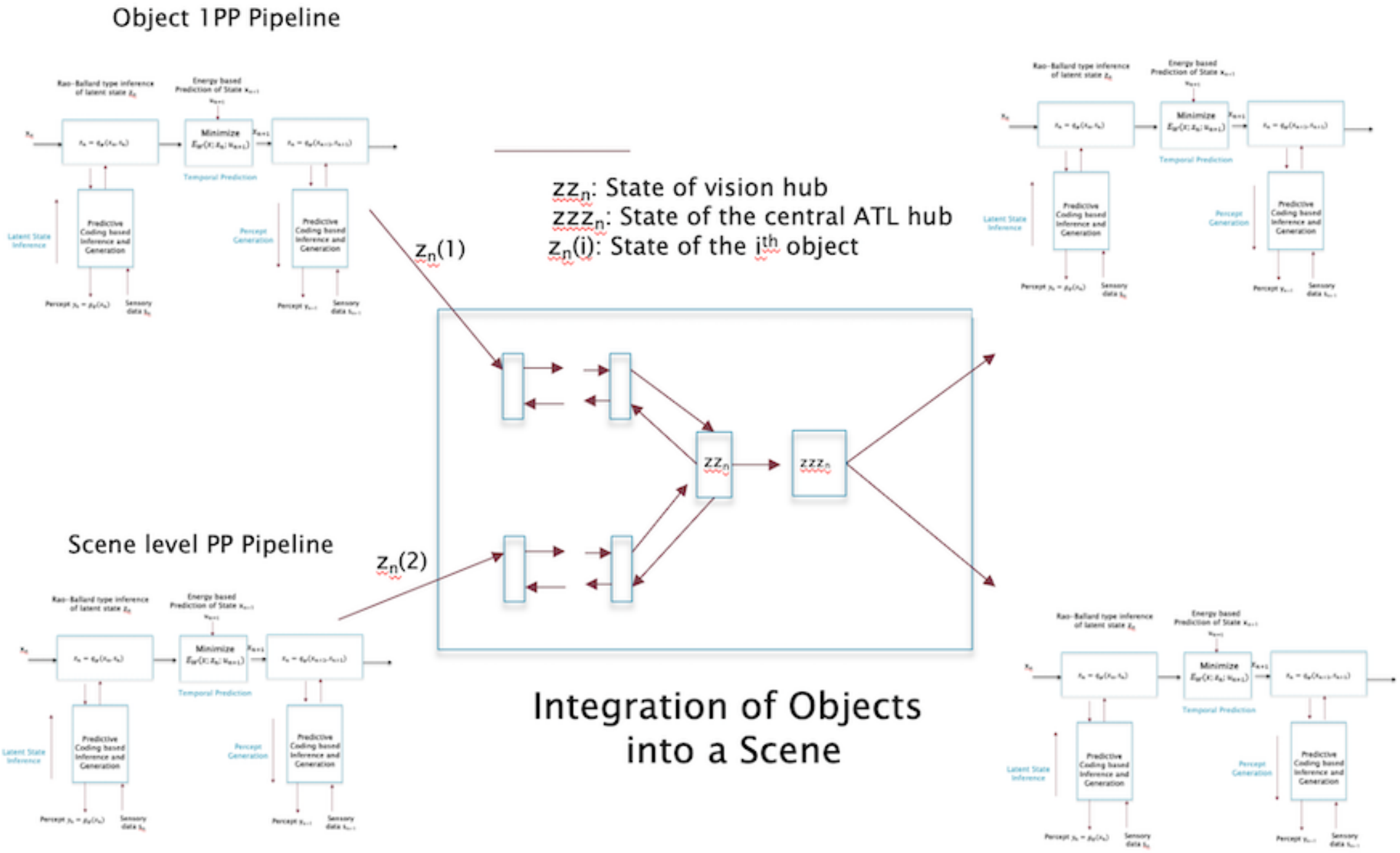

Figure 6: Integration of Multiple Object and Scene Level Predictive Processing Pipelines into the Vision Hub at Level 2

The structure of the level 2 vision hub is shown in the above figure. It shows that the latent state $z_n$ from the object level pipeline is fed into an object level predictive coding pipeline in the vision hub, where $z_n$ serves as the ground truth value. As a result of this architecture the hub latent state $zz_n$ gets modified, and the new value reflects the value of the latent state $z_n$ at the object level. The presence of other predictive coding pipelines at the hub ensures that the hub state $zz_n$ reflects not just the latest information from the current object level pipeline, but also information from all the other object pipelines that are active at the same time as well as information from the always-on scene level pipeline. Thus $zz_n$ serves as an integrated representation for the vision system. If an object moves out of the field of vision, its level 3 predictive processing pipeline is dis-connected from the level 2 vision hub after a time lag, and as other objects appear their pipelines are in turn connected to the hub. If the objects are well known due to frequent appearance, then their predictive processing models are pre-trained and stored in memory, while new objects undergo a period of training.

When there are several objects in the field of vision, our attention focuses on only one at a time, and the predictive processing pipeline for its model is updated with fresh sensory data. The IM-LEPP model proposes that the state for the other active objects also get updated, but that happens in an open-loop fashion without sensory data that was described in Varma, based on the last known position and movement of the object.

Note that the structure of the predictive coding pipeline agrees with what has been observed about the variation in the mode representation in the biological ATL hub as shown in figure 2. Specifically the predictive coding states that are closer to the vision (or language signal) mode still retain the signatures of that mode, and as we move up the pipeline, the representation turns gradually amodal.

This design takes into account the predictions that arise when two objects are interacting with each other, for example a ball bouncing off a wall. In this case there is a level 3 pipeline for the ball, as well as for the wall, and their latent states are integrated at the level 2 hub, and subsequently this information is used to update the energy function $E_W(x; z_n, u_{n+1} = zzz_n)$ for the ball. Thus if the ball is very close to the wall, then the next prediction will lead to a change in its trajectory since the ball model knows about the presence of the wall.

If both the objects are moving towards each other, for example two balls about to collide, and assume that our attention is focused on ball 1 so that its state is being updated constantly with fresh sensory data. In this situation the model's level 3 predictions for ball 2 operate in an open-loop fashion based on its last position and velocity, and continue to get integrated with ball 1's predictions at the level 2 hub. Thus the system should be able to predict the collision at the right instant even though attention is focused on ball 1, provided the hub's integration cycle is fast relative to how quickly the two objects' relative positions are changing.

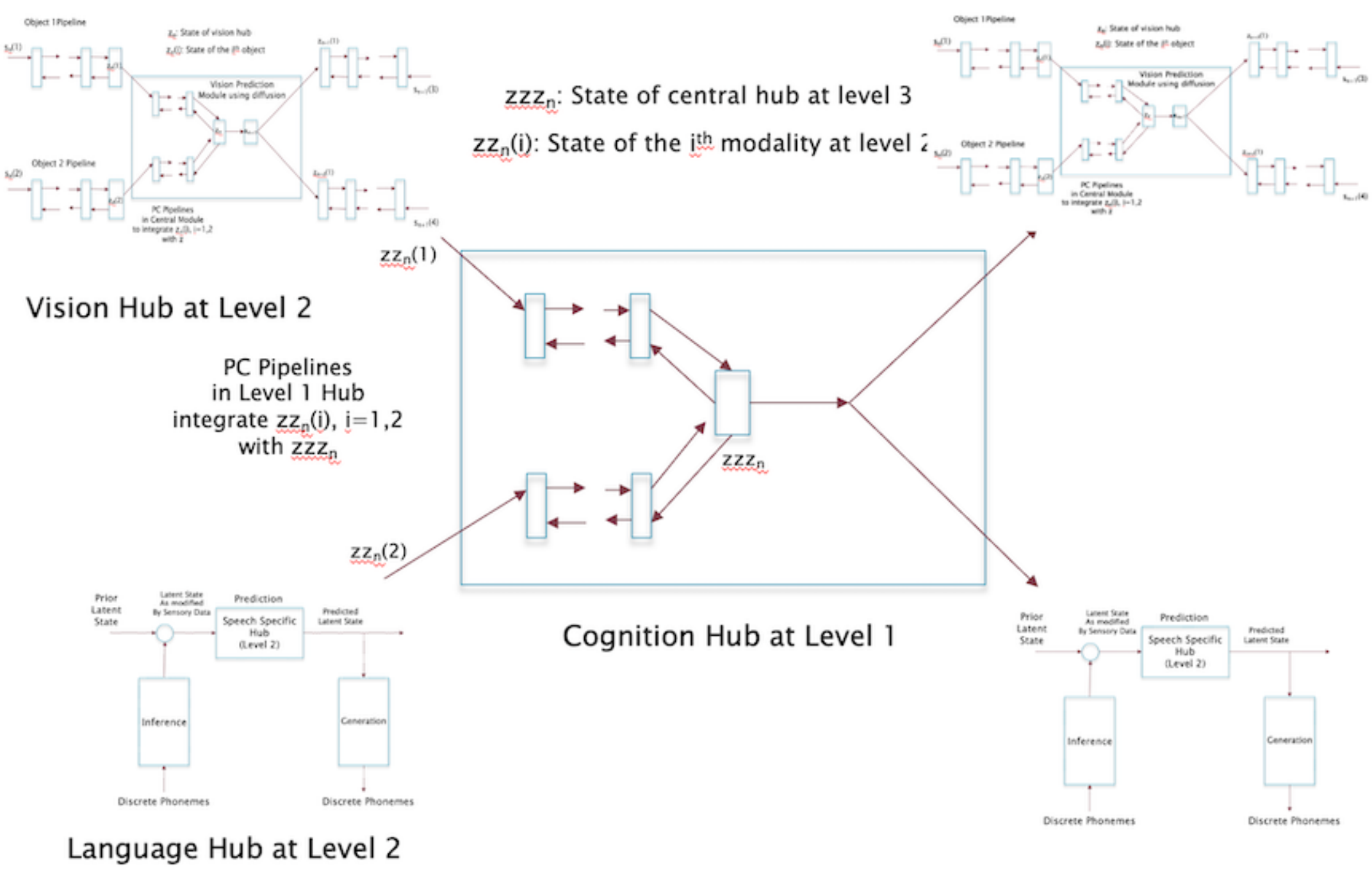

Figure 7: Integration of Vision and Language Hubs into the Cognition Level Hub at Level 1

The above figure shows the structure of the level 1 hub located in the ATL. It is the same design as for the level 2 vision hub, however now the signals $zz_n(1)$ and $zz_n(2)$ are coming in from vision as well as language modalities. There is a predictive coding pipeline for vision and another one for language within the hub, and this results in a representation $zzz_n$ that is amodal and is influenced by both. After the predictive coding pipeline settles down, the resulting value of the hub state $zzz_n$ is fed back into the vision and language pipelines, and serves as a conditioning variable in the energy function $E_W(x; z_n, u_{n+1} = zzz_n)$ used to generate the prediction $x_{n+1}$ which is used to generate the next percept $g_\psi(x_{n+1})$. Thus the percept generation takes place at the individual mode level, but takes into account everything else that is happening by virtue of this architecture. The latent $x_{n+1}$ subsequently gets modified by the new sensory data $s_{n+1}$, which results in the latent state $z_{n+1}$ and the cycle continues.

Note that the vision updates happen almost continuously, while the language updates in the form of words happen every hundreds of milliseconds. The problem of integrating two streams running on different time scales is called the *multirate sensor fusion problem* in robotics. The proposed solution to this problem as shown in the above figure has empirical support in psycholinguistics. the Visual World Paradigm literature Tanenhaus, Spivey-Knowlton, Eberhard, & Sedivy, 1995 showed via eye tracking that visual context influences spoken word recognition and syntactic parsing during the earliest moments of processing. We will assume that the update to the shared state $zzz_n$ on the arrival of a new word happens much faster than the interval between successive vision updates, in other words the system does not run into the problem that the target for the word update is continuously changing as a result of the faster vision updates.

Consider the scenario in which we are driving a vehicle and there is another vehicle in front of us, as well as other objects such as pedestrians etc which are visible. As per the model, the system will install level 3 predictive processing for all objects that we pay attention to during the course of the drive, but there are some objects that we pay more frequent attention to, such as the car in front of us. The scenario can evolve into one of the following ways:

- We are paying frequent attention to the car in front of us, say car 1, so its representation $z_n$ gets updated frequently with new vision data. There is another car behind us that we can see through our rear view mirror, say car 2, but we play less frequent attention to it. In between the times that we look into the rear view mirror, the representation for that car 2 evolves as expected, so that if that car appears by the side of our vehicle it is something that the system has predicted (i.e., the latest level 3 prediction $g_\psi(x_n)$ and the sensory data $s_n$ for car 2 when it appears would agree with each other), so we are not surprised. The actual mechanism that the system uses for tracking the behavior of car 2 is open loop prediction using a diffusion model from the last known position of the car. This mechanism was described in Varma.
- Consider the same scenario as before, but assume that car 2 behind us made a turn while we were not paying attention to it. In this case our visual model does not know that the car has disappeared, and the next time we look back we expect to see it, but the sensory data says otherwise. Hence car 2 is not behaving in the way that our internal model is expecting it to behave, which causes some surprise. Once we see that car 2 is not there, its level 3 pipeline is removed from the set of objects that are being tracked at level 2. Another example of this would be if car 1 in front of us suddenly braked while we were looking at our phone. Again in this case our internal model for car 1 is updating in a way that does not agree with reality, until we look up and see the change. Unlike the case for the disappearing car, in this case the model for the other vehicle is not de-installed, but is updated with a large error correction when new sensory data comes in, i.e., when we look up from our phone.
- There are cases in which the brain actively suppresses the updating of one or more objects whose pipelines are currently active. For example while driving we don't want to get distracted by the image of large billboard by the side of the road. In this case a level 3 pipeline will be installed for the billboard when we see it for the first time, however we decide not to pay attention to it. This can be done by reducing the value of the precision parameter $\Pi_y$ that is used in the predictive coding pipeline for the billboard at the level 2 vision hub. This mechanism was first described by Feldman and Friston (2010) in their paper *Attention, uncertainty and free energy*. This mechanism for reducing the value of $\Pi_y$ can also be invoked to gradually reduce the effect of automatic updates coming from objects to whom we haven't paid attention in a long time.
- As the last example, consider the case when a level 3 pipeline for an object that is in the vicinity of our car never gets installed, so that we are completely unaware of its existence. This corresponds to the case when there is a car in our blind spot, and we try to make a turn into its lane. In this case there is a much bigger surprise waiting for us when the other car honks, and at that point the system installs a model for it and starts tracking it. Another well known example of this phenomenon is the case of bear costumed person crossing the court in the middle of a basketball game and it fails to get noticed, as was originally pointed out by Simons and Chabris. Again since attention is focused on the ball and the players, the level 3 model for the bear never gets installed.

The last scenario points to the need for a cheap always-on detector, separate from the pipelines it might spawn, since something has to decide when to instantiate a new level-3 pipeline, and that decision process itself must be resource limited. Itti, Koch and Niebur's (1998) saliency-map model is the classic computational proposal. They proposed a cheap, bottom-up, parallel process computing simple feature contrasts (color, intensity, orientation, motion) across the entire visual field continuously, producing a topographical saliency map that competes to determine where attention deploys next. A new level 3 pipeline gets instantiated only when a region's saliency signal is large enough and wins the competition for currently available attention. In the bear case, task absorbed attention means the saliency signal from the costumed figure never wins that competition, so no pipeline ever gets created. Note that this is a separate, lighter-weight mechanism than the always-on scene pipeline described above, since it is a gating signal rather than a full predictive-coding pipeline.

This is consistent with inattentional-blindness follow-up work showing measurable physiological orienting responses to "unnoticed" stimuli despite no conscious report. This gives a concrete mechanism for the "background things vs. stuff" distinction, everything starts as coarse "stuff" background by default and a region gets promoted to its own individuated "thing" pipeline specifically when its saliency signal wins the attention competition. But the fact that an object has been given its own pipeline does not mean that we are constantly monitoring it, which can result in occasional surprises when it does not behave as expected.

## 4.1 Perceptual Multistability: The Necker Cube

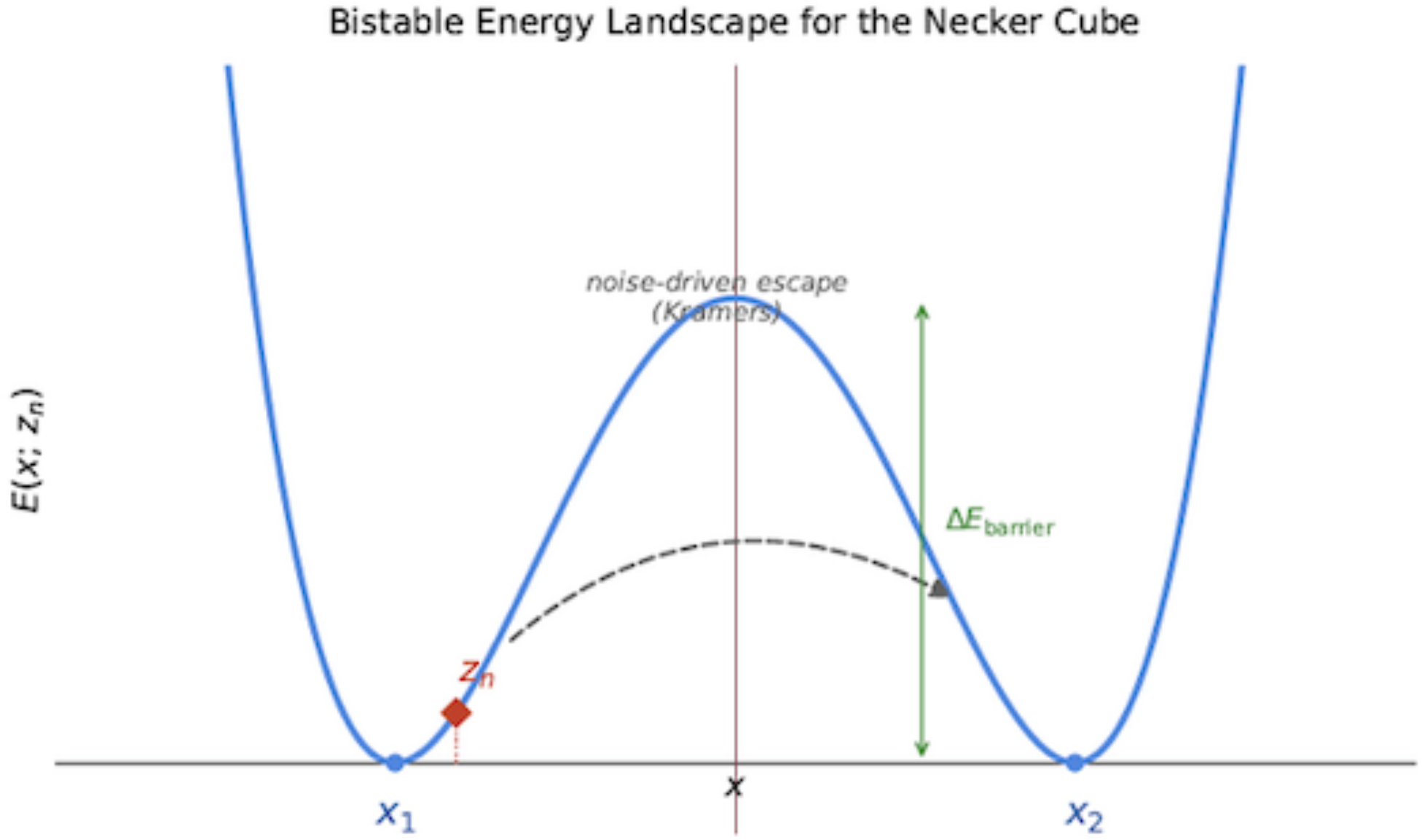

Figure 8: Energy Landscape for the Necker Cube

A useful test case for the level 3 energy landscape used for predictions, is a genuinely ambiguous stimulus such as the Necker cube, whose line drawing is equally consistent with two distinct three-dimensional interpretations. Let $x_1$ and $x_2$ denote the two corresponding minima of the object's prediction energy $E(x; z_n)$ as shown in the above figure. Both minima are shaped by the same fixed sensory input $s_n$ (a two-dimensional wireframe that does not itself resolve depth), so the landscape's overall two-well structure is a property of the stimulus itself, present regardless of which interpretation is currently held.

Suppose the system currently occupies $x_n = x_1$ as the current prediction. On the next cycle, new sensory data arrives, and the correction step $z_n = q_\phi(x_1, s_n)$ combines this prediction with the incoming (still ambiguous) evidence. Critically, because $s_n$ is genuinely uninformative between the two interpretations, the prior term should be weighted heavily relative to the sensory term at this level, which is the same precision-weighting mechanism already used elsewhere in this paper (e.g., the billboard-suppression case). Thus $z_n$ remains close to $x_1$ rather than being pulled toward some interpretation-neutral point as shown in the figure. The subsequent prediction is then obtained by minimizing $E(x; z_n)$, a landscape retaining the same two wells but now with the starting state asymmetrically situated near the previous prediction $x_1$ (see above figure).

Because the sensory evidence never favors one interpretation over the other, there is no deterministic force in this energy minimization that ever favors leaving the current well, In other words, absent noise, the system would settle into whichever interpretation it first reached and remain there indefinitely. Perceptual switching is therefore possible only through the stochastic component of the diffusion-based optimization which can cause an escape over the energy barrier separating $x_1$ and $x_2$, in the manner described by Kramer's (1940) escape rate,

$$rate \propto e^{-\frac{\Delta_{barrier}}{T}}$$

rather than through any evidence-driven reshaping of the landscape itself.

This distinguishes Necker cube switching from the garden-path reanalysis discussed in the language section in a later section. Language reanalysis is driven by the arrival of new, disambiguating evidence that reshapes the landscape (Figure 8 in Varma), while Necker cube switching occurs under a landscape that remains fixed, driven by the diffusion process's own persistent noise term. The two phenomena therefore require the same general architecture, which is a bistable energy landscape traversed by stochastic optimization, but differ in whether it is the landscape or the occupant that changes.

This picture also makes a specific, checkable prediction about the statistics of switching. Because each new prediction cycle is warm-started from the currently-held state rather than independently resampled, successive predictions are correlated rather than independent draws. Thus dominance durations should follow a distribution with a mode away from zero and positive serial correlation between consecutive durations, rather than the memoryless (geometric) switching that independent resampling would produce. This is consistent with the pattern generally reported for bistable perception.

This analysis further predicts that a stimulus with an asymmetric prior, such as the well-documented "viewing-from-above" bias in Necker cube perception, whereby observers show a persistent preference for interpreting the figure as viewed from above, attributed to the greater ecological frequency of looking down at objects than up at them (Wernery, Atmanspacher, & Kornmeier, (2015)) should correspond to an asymmetric energy landscape in which, say the orientation one is favored and thus has a deeper minima $x_1$, which causes the $\Delta E_{barrier}$ to be deeper as well, leading to the observed asymmetry. Given the exponential sensitivity of the Kramers rate to barrier height, this predicts systematically longer dwell times in the preferred interpretation, exactly as observed.

## 5 A Predictive Processing Model for Language

Language comprehension and generation is a relatively late addition to our cognitive system, it happened in the last 50,000 years. By then the brain had already built a rich, grounded, pre-linguistic continuous conceptual state (the $zzz_n$ in the IM-LEPP model), and language was added comparatively effortful, serial auto-regressive-like process that reads from and writes to that substrate rather than building another one from scratch. Hence the brain's $zzz_n$ equivalent at the central ATL hub is grounded in perception, sound, social experience etc and language is another channel feeding into this already existing representation. This points out to a fundamental difference between the IM-LEPP model and LLMs, since the latter are not grounded and base their latent representations purely from word co-occurrence statistics and nothing else. There have been some recent efforts to build a predictive processing model for language, for example see the LD4LG model. However this and similar models run into the problem of designing a good auto-encoder for language, which hasn't been solved yet. The reason for this is that their latent representations are not grounded with other sensory data, and no amount of auto-encoder engineering can fix lack of grounding.

Language shares a number of features in common with the visual system, and both can be modeled as predictive systems, images in one case and words in the other. That said, words are a more abstract entity than objects and are discrete in nature, i.e., there are a finite number of them. The problem then is that of translating the sound coming in through our ears or patterns coming through our visual system, into word level latent representations that attach semantic meaning that can be used for prediction of the next word. Language signals could be coming from an external source, or it could be from ourselves in the form of speech or writing. In the former case, the next word in the sequence can serve as an error connection signal for the brain's prediction for the next word, very much like the error correction that happens in vision, and this can be used to train the brain's model for next word prediction. For the case when we are the originators of the signal, there is still an error correction feedback loop in operation, since sometimes we speak out a word, and then hearing the sound makes us realize that we meant to say something else (or when we write a word and then revise it).

There has been a good amount of evidence that has been collected over the years that unlike for vision, the brain uses discrete representations when processing language. The paper on categorical perception by Liberman, Harris, Hoffman & Griffith, (1957) is a classic in this area. They showed that a continuously-varying acoustic parameter is perceived and discriminated categorically, with a sharp identification boundary. Direct neurophysiological confirmation was established by Chang et al. (2010) who found categorical speech representations in human superior temporal gyrus (STG). Mesgarani, Cheung, Johnson & Chang (2014), using intracranial recordings, found STG populations tuned to discrete phonetic features, not raw continuous acoustics.

So the brain does discretize but not at the level of the word vocabulary size, which easily exceeds 10,000 words for most people. Instead, a hierarchical cascade of much smaller categorical decisions are made at the phoneme level on a small set of articulatory-feature dimensions (voicing, place, manner, and a few others) identified by Mesgarani et al., and then several phoneme representations are strung together to compose into a word representation. Hence while phoneme level feature encoding is well established (see Leonard et al. (2024)), exactly how these compose into whole-word representations is still an open question. In this paper we propose a hierarchical two level mechanism by which this can be accomplished.

We will start with the simpler case first when the source of language is through our visual system, i.e. by reading or writing.

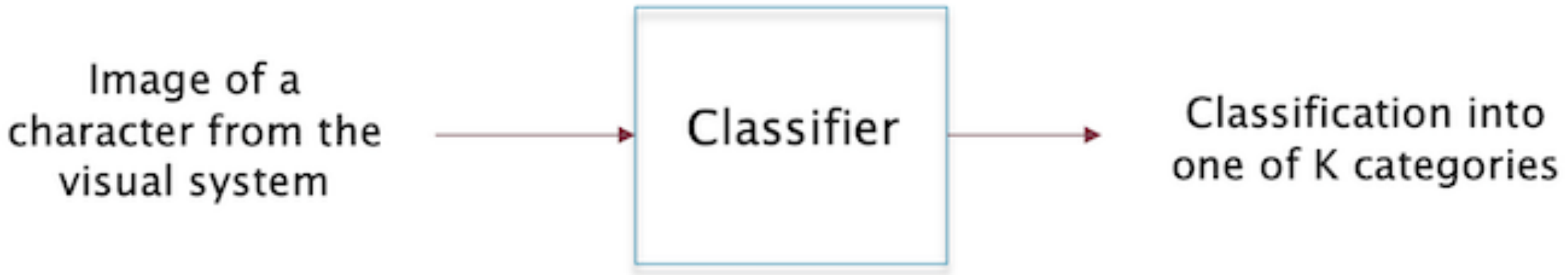

Figure 9: Classification of a character image into one of K categories

Consider the simple case of child reading, in which case he or she reads one character at a time. Adults on the other hand use a faster system in which multiple characters are read at time in order to speed up the process, which comes after practice. We will assume that the image of an individual character coming in from the visual system is sent to a classifier, that outputs a discrete number indicating the category. There can be 26-70 different categories depending upon the punctuation. This process results in a discrete sequence $(ch_1, \ldots, ch_N)$ that is fed into a predictive processing pipeline.

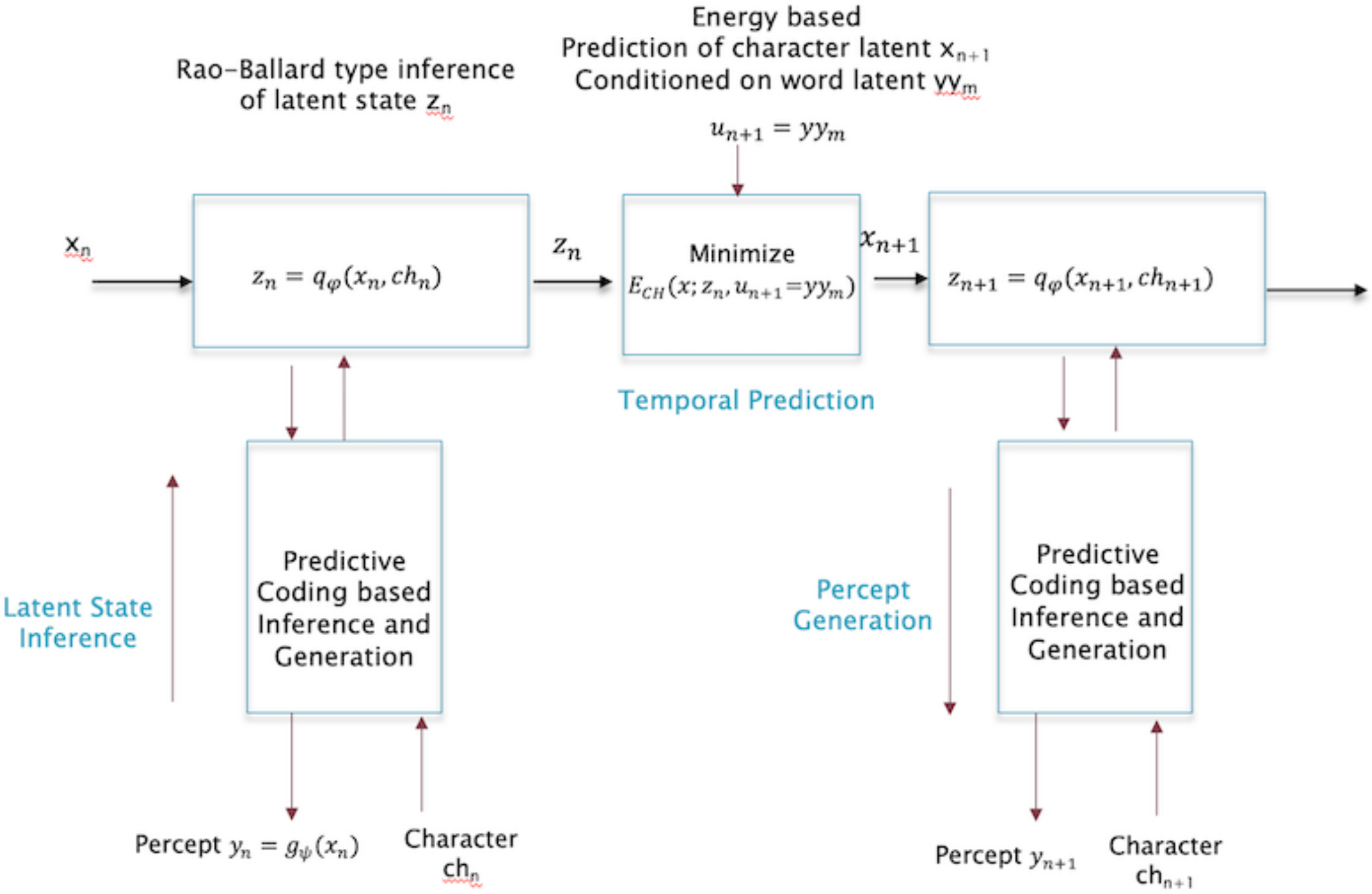

Figure 10: Predictive Processing Pipeline for Next Character Prediction

The discrete character sequence is sent into a predictive processing pipeline that creates an latent representation $z_n$ for each character using the predictive coding function $q_\phi$, followed by prediction of the next latent representation $x_{n+1}$ using a energy based diffusion model $E_{CH}(x; z_n, u_{n+1} = yy_m)$. Note that this prediction is conditioned on not just the previous character level latent $z_n$, but also on a latent representation $yy_m$ for the $m^{th}$ word (whose characters are being generated), which comes from the level 2 word pipeline. The computation of $yy_m$ is explained below when we discuss the word level pipeline. Next $x_{n+1}$ is converted into a character using the predictive coding function $g_\psi$. The generated character $y_{n+1} = g_\psi(x_{n+1})$ is compared with the actual next character $ch_{n+1}$, and the error generated by this information is used to estimate the next representation $z_{n+1}$. These operations are illustrated in the above figure.

The system is able to detect the end of a word by tracking the difference between the predicted latent state $x_{n+1}$ and the actual latent state $z_{n+1}$. If this difference exceeds some threshold, then the model assumes that the previous character $ch_n$ occurred at the end of a word (saw the $m^{th}$ word $w_m$), and the representation $z_n$ is sent over to the word level pipeline as discussed next. We also use the empty space at the end of each word to infer the word ending, however there some languages such as Chinese and Old Latin that don't use spaces (see Saenger 1997), and the alternative mechanism described here is more general, and also works for the case of auditory signals.

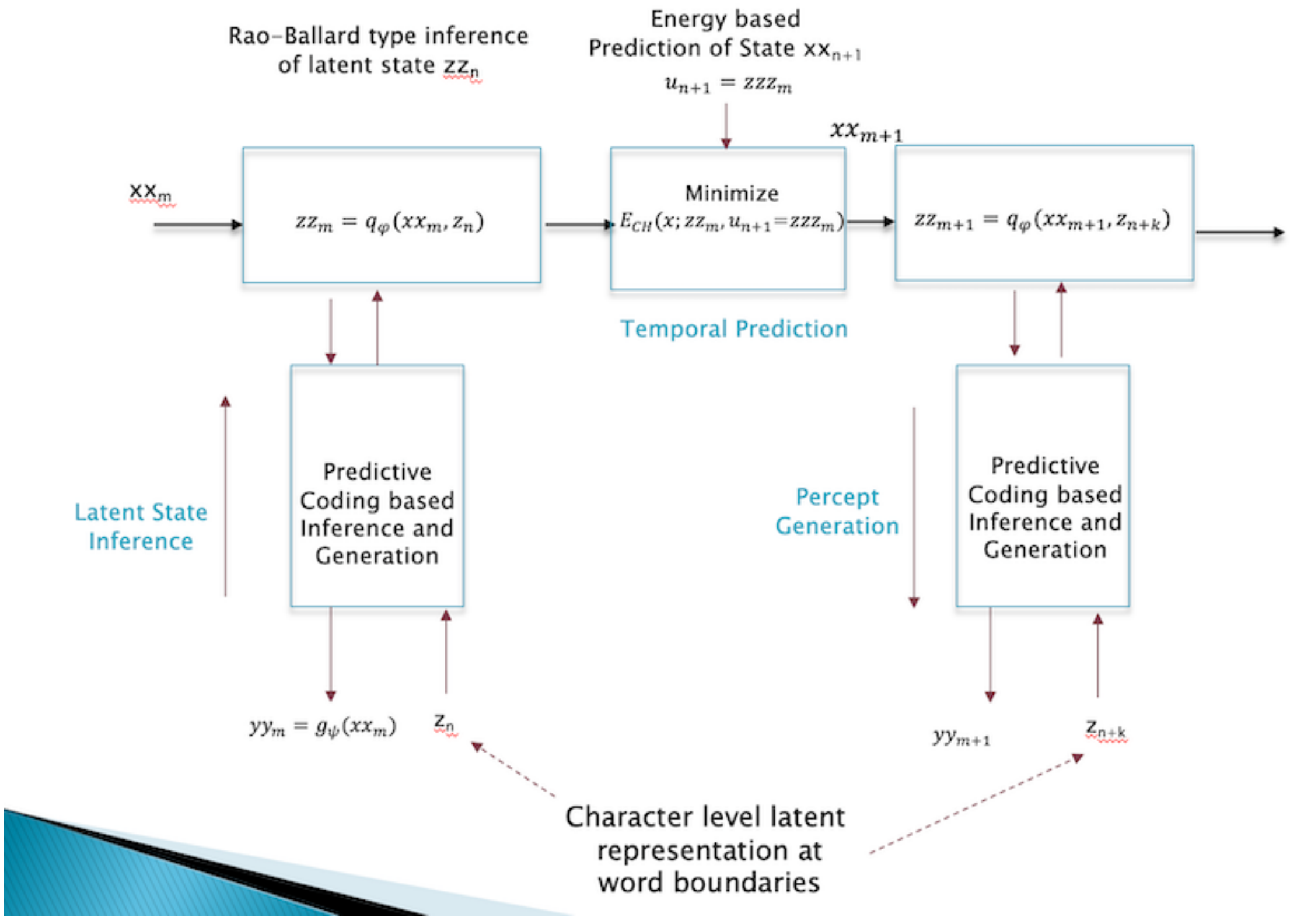

Figure 11: Predictive Processing Pipeline for Predicting the Latent Representation for the Next Word

The predictive coding pipeline at the word level uses the latent state $z_n$ from the character pipeline as the ground truth that represents the latent representation for the $m^{th}$ word. Note that the word level subscript $m$ for the current word is not the same as the character level subscript $n$ for obvious reasons. This pipeline creates a higher level latent word representation $zz_m$ by modifying the existing representation $xx_m$ to $q_\phi(xx_m, z_n)$. $zz_m$ is then sent to the level 1 central ATL hub where it gets modified by the vision data to the latent $zzz_m$. For example if the current image is that of an apple, then this is reflected in $zzz_m$.

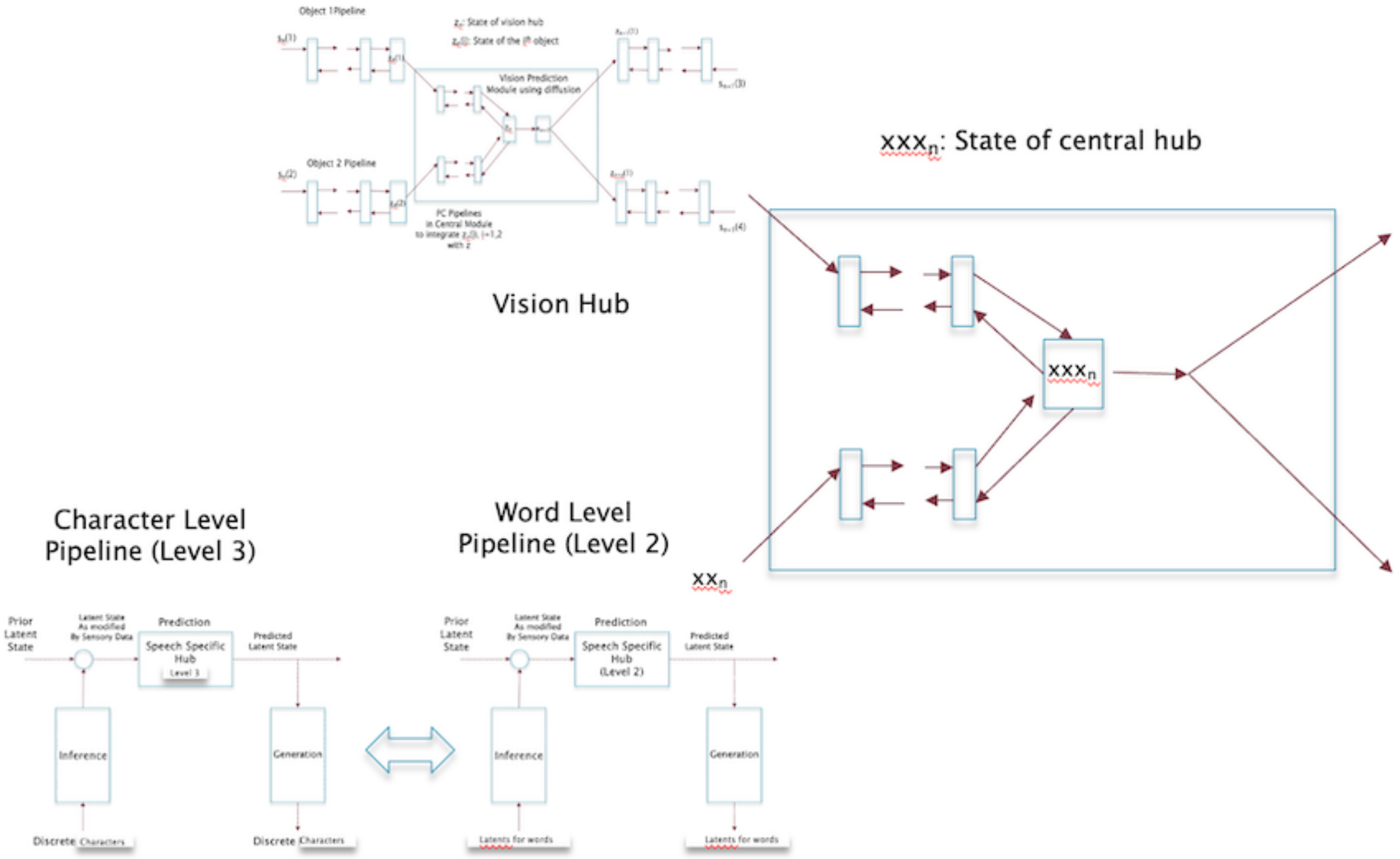

Figure 12: Integration of vision and language modules at the central ATL hub

The latent $zzz_m$ is then fed back to the level 2 word pipeline where it is used to predict the next word by using the the energy based prediction module $E_W(x; zz_m, zzz_m)$ and this results in the prediction $xx_{m+1}$ for the next word latent (see above figure). Note that, unlike the vision hub at level 2, which performs integration only, the word-level pipeline includes its own dedicated prediction step. This reflects the fact that phoneme-level prediction serves segmentation, while word-level prediction serves ordinary sentence-level anticipation, a genuinely distinct function operating at a different timescale. This mechanism also illustrates why the IM-LEPP model may be able to learn new words faster, since the predicted word is not only a function of the previous word level context $zz_m$, but is also influenced by the vision modality by means of the latent $zzz_m$. Similarly the emotional related information coming in through the valence system (valence) can influence our choice of the next word.

$xx_{m+1}$ is subsequently used to generate the next word latent $yy_{m+1} = g_\psi(xx_{m+1})$, and this value is fed back into the level 3 character level predictive processing pipeline shown in figure 10, where it influences the prediction of the next character $x_{n+2}$ through the energy function $E_{CH}(x; z_{n+1}, yy_{m+1})$. This in turn gets modified by the other characters in the next word being read. When the last character of that word is encountered, then the latent representation at that time $z_{n+k}$ (where $k$ is the number of characters in the word just read) is fed back into the word level model to correct the prediction $yy_{m+1}$, and this closes the word level prediction loop. As pointed out in the introduction, this is also an hierarchical system, but the hierarchy is in time rather than in space, as was the case for the visual system.

For the case when we are doing character generation, i.e., writing, this feedback loop between the character level and word level pipelines is still active. In this case it serves as a verification of whether the word that was generated at level 3 matches the word that the level 2 word level system meant to generate.

The hierarchical system composed of a discrete character (or phoneme) level pipeline at level 1 with a continuous word level pipeline at level 2 constitutes a possible solution to the whole word representation problem posed in Leonard.

The three level model for language that has been presented here is backed up by some experimental data that has been collected over the years. Hickok and Poeppel's (2007) dual-stream model locates early phonological processing in STG/STS of the brain with a lexical-interface stage in posterior middle temporal gyrus (pMTG). This maps cleanly onto the three stages of the model's auditory pathway (developed in full below), namely STG/STS for phonetic features, pMTG for word-level representation, and, continuing upward, Lambon Ralph's ATL hub for integration with other modalities.

Do the latent states $zzz_m$ or $zz_m$ encode 'thought'? Levelt's production model as described in his book Speaking: From Intention to Articulation (1989) has a first stage, conceptualization, whose output is a pre-verbal message i.e., a language-independent conceptual representation of what to say, prior to and dissociable from any particular verbalization (which is why "the same thought" can be expressed in different words or languages). This is a direct architectural instantiation of exactly the proposed role for $zzz_m$ as a persistent, amodal state that the generative pathway then unrolls into a word sequence, with $zzz_m$ playing the role of the preverbal message and the language pipeline playing Levelt's formulation stage.

In order to see the see multi-modal language model in operation, consider the following scenario: The model is initially shown a video, which results in the invocation of the vision model. Lets assume that the ATL hub state at the end of the video is $ZZZ_1$. Then the user asks the model a written question that has to do with the video, and this information comes in through the language model, and at the end of the question the ATL hub state is at $ZZZ_2$. Note that this hub state is now a summary of all that the model has seen and read so far, and is now primed to provide an answer. This it does by feeding $ZZZ_2$ back into the level 2 next word model, which results in the production of characters at the level 3. These characters are then fed back into the level 2 word model (i.e. the model self reads to make sure that the characters are consistent with the predicted word), and this results in the prediction of the next word, and an update to the word model state $zz_m$. The updated state is then fed into the ATL hub, so that it gets updated with any new information that has come in through any of the other modalities, and then gets fed back to the word model to generate the next word. This cycle continues until the model has generated the answer to the question.

Modern multimodal LLMs can also be used for the scenario described above, but there is an important distinction between the way the IM-LEPP model operates compared with the LLM. The LLM presumably has the equivalent of an internal latent representation *zzz* that summarizes everything that model has seen, read and generated so far. However this latent state is not directly observable, though experiments done by probing the middle layers of the transformer have been shown to have characteristics that are similar to *zzz*. In contrast, the latent state in the IM-LEPP model is explicitly made available, at all levels of operation. This allows us to directly probe the 'mind' of the model, and also influence it through the ATL hub modalities. For example one can imagine creating an emotion state for the model by tracking how well its latent state predictions match with sensory data thus simulating the amygdala. The state of the artificial amygdala can in turn be fed back into the ATL hub to influence the model's central latent state, and thus its future word or image generation. Note that this type of operation cannot be done with LLMs due to lack of access to its latent state.

### 5.1 The Predictive Coding Pipeline for Discrete Sensory Data

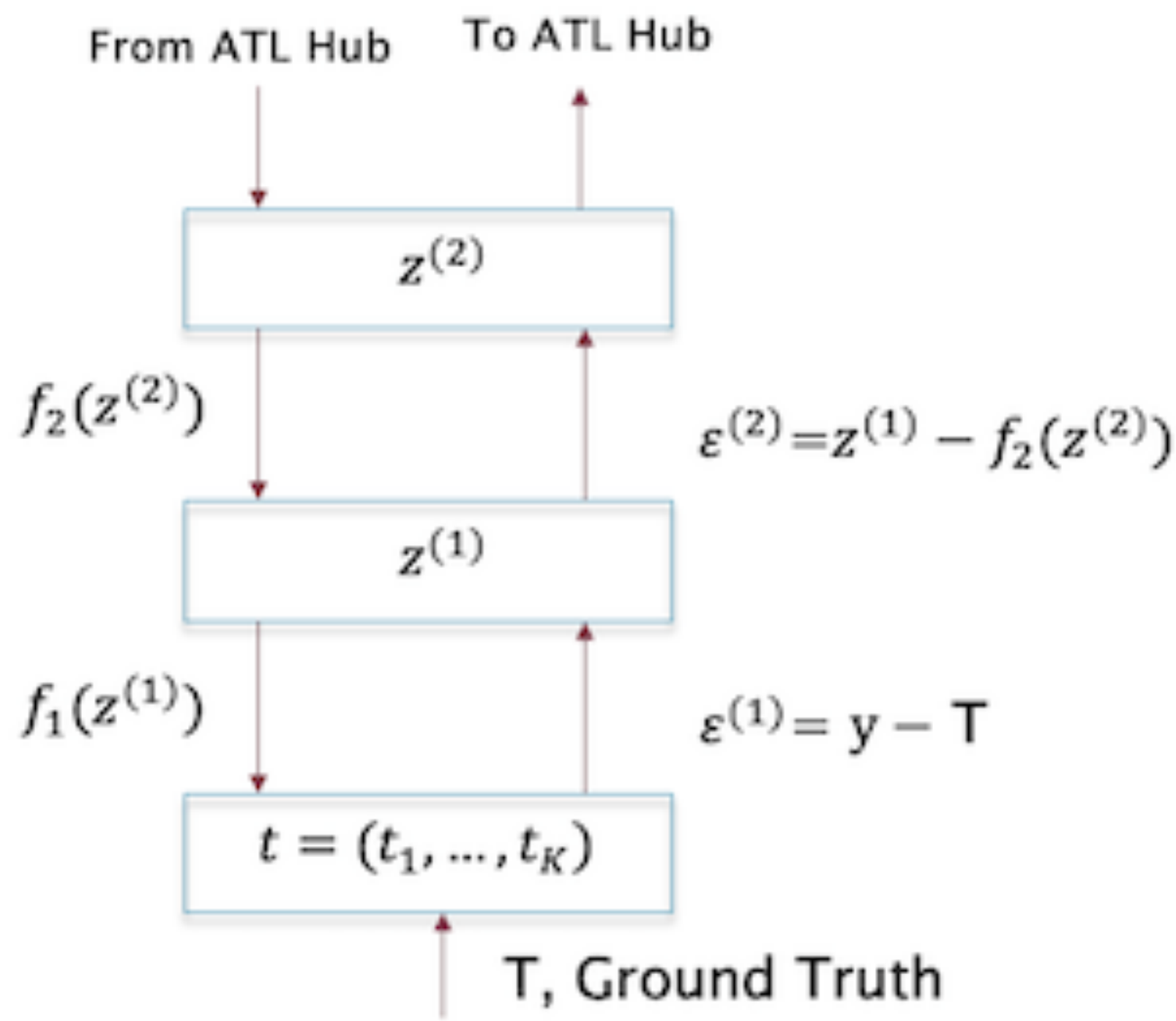

t can take one of K discrete values (1,0,...,0),(0,1,...,0),...,(0,0,...,K) with probabilities

$$p(t = (t_1, \dots, t_K)) = (y_1)^{t_1} \dots (y_K)^{t_K} \quad \text{where } y_k = \frac{e^{z_k^{(1)}}}{\sum_i e^{z_i^{(1)}}}$$

Problem: Choose $z^{(1)}$ so that it better predicts $t$ and also stays close to the prediction $f_2(z^{(2)})$ from the higher layer

Figure 13: Predictive Coding Pipeline for Characters

The traditional predictive coding pipeline was designed for analog sensory data, but it can be extended to the case when the data happens to be discrete, as for the case of characters or phonemes in the level 3 part of the language model (see Whittington and Bogacz). The above figure shows the computational details for the predictive coding model at the character level. The system processes characters one at a time, and for each character it builds up an internal representation using predictive coding. We will assume a simple two level hierarchy, though the model allows for any number of levels. The top level has a latent representation $z^{(2)}$, and this used to generate a representation $z^{(1)}$ at the lower level using a function $f_2(z^{(2)})$ and this representation in turn generates the final representation $t = (t_1, \ldots, t_K)$ where $K$ is the number of characters being modeled. The representations $z^{(1)}$ and $z^{(2)}$ are in continuous space, however $t$ lies in discrete space and uses the 1-hot representation so that individual characters $(ch_1, ch_2, \ldots, ch_K)$ are represented by $(1,0,\ldots,0), (0,1,\ldots,0), \ldots, (0,0,\ldots,1)$ respectively. The output $t$ is generated from $z^{(1)}$ by a process of sampling using the distribution

$$p(t = (t_1, \ldots, t_K)) = (y_1)^{t_1}(y_2)^{t_2} \cdots (y_K)^{t_K}$$

where $y_k$ is given by the Boltzmann distribution (also called the softmax function in machine learning)

$$y_k = \frac{e^{z_k^{(1)}}}{\sum_i e^{z_i^{(1)}}}$$

In this equation $z_i^{(1)}$ is the $i^{th}$ component of the vector $z^{(1)}$. Lets assume that $z^{(1)}$ leads to the probabilities $y = (y_1, y_2, \ldots, y_K)$ while the ground truth is given by $T = (T_1, T_2, \ldots, T_K)$ This generates an error $\epsilon^{(1)} = y - T$, which is propagated to the level above. Using the Bayesian argument used in predictive coding, it can be shown that optimal $z_i^{(1)}$ is obtained by minimizing the energy function

$$E_{CH}(1) = -\sum T_i \log y_i + \frac{1}{2}\epsilon_{z^{(1)}}^T \Pi_1 \epsilon_{z^{(1)}}$$

where $\epsilon_{z^{(1)}} = z^{(1)} - f_2(z^{(2)})$. Using gradient descent $z_i^{(1)}$ is updated according to

$$z_i^{(1)} \leftarrow z_i^{(1)} - \eta\left[(y_i - T_i) + \Pi_1 \epsilon_{z^{(1)}}\right]$$

Note that all the information required to update $z_i^{(1)}$ is available locally.

### 5.2 Extension to Auditory Processing of Language (Phonemes)

The predictive coding model for level 3 developed above assumes a single flat categorical readout, appropriate when the input alphabet has no independently-motivated internal structure, which is the case for written characters. Auditory input requires a modification, because phonemes are not atomic: articulatory phonology decomposes each phoneme into a small set of independent distinctive features (voicing, place of articulation, manner of articulation, nasality, and a small number of others), and direct intracranial recordings from human superior temporal gyrus confirm that this is the level at which the auditory cortex actually represents speech sound. Thus there are neural populations that are tuned to specific feature values, and not to whole phonemes as unitary categories (see Mesgarani, Cheung, Johnson, & Chang, 2014).

We accommodate this by replacing the single $K$-way categorical readout with $D$ independent, smaller categorical readouts, one per feature dimension $d = 1, \ldots, D$ (typically $D \approx 6-8$; e.g. $K_{\text{voice}} = 2$, $K_{\text{place}} \approx 7-8$, $K_{\text{manner}} \approx 6-7$). The latent $z^{(1)}$ is partitioned into $D$ corresponding slices, $z^{(1)} = (z^{(1,1)}, \ldots, z^{(1,D)})$, and the emission energy becomes a sum of independent cross-entropy terms:

$$E_{PC}(1) = \sum_{d=1}^{D} \left[ -\sum_i T_i^{(d)} \log y_i^{(d)} \right] + \frac{1}{2} \epsilon_{z^{(1)}}^T \Pi_1 \epsilon_{z^{(1)}}, \qquad y^{(d)} = \text{softmax}\big(z^{(1,d)}\big)$$

This is justified by treating the $D$ feature-level classifiers as independent experts whose distributions combine multiplicatively (Hinton, 2002). In log space this is exactly a sum of energies, so the joint model remains a single, well-defined energy function despite the factored readout. The resulting gradient update is unchanged in form from the single-category case, and remains fully local: each slice updates using only its own residual and its own portion of the prior term,

$$z^{(1,d)} \leftarrow z^{(1,d)} - \eta \Big[ \big(y^{(d)} - T^{(d)}\big) + \Pi_1^{(d)} \epsilon_{z^{(1,d)}} \Big], \qquad d = 1, \ldots, D$$

with no cross-feature terms required in the bottom-up direction.

The independence assumption is reasonable for inference since each feature can be estimated from the acoustic signal largely on its own, but this does not hold for the top-down prior. Phonemes occupy only a small, structured subset of the full combinatorial space of feature-value combinations and most combinations correspond to no phoneme in any language. The generative map $f_2(z^{(2)})$ must therefore produce a *jointly* consistent prediction across all $D$ slices simultaneously, encoding the correlations between features that define the phoneme inventory, rather than predicting each feature independently. Absent this, the model would treat phonologically illicit feature combinations as being just as expected as licit ones.

This modification is confined entirely to the level-1 emission layer. The temporal prediction module is unaffected: it continues to operate on a single continuous latent and to output a continuous prediction $x_{n+1}$ via the diffusion process described above, which is then passed through the factored readout described here rather than through the single flat softmax used for the character case.

### 5.3 Approximating the Softmax Function in Brain Circuitry

The computations in the predictive coding pipeline described above require the computation of $y_i$ using the softmax function. How can this be accomplished in the brain?

A substantial body of work by Carandini and Heeger identifies **divisive normalization** as what they term a canonical neural computation, i.e., as a a single computational motif recurring, with only minor variation, across an unusually wide range of brain systems and species (Carandini & Heeger, 2012). In its general form, the response of an individual neuron is divided by a term reflecting the pooled, summed activity of a local population of neighboring neurons:

$$response_i = \frac{R_i^n}{\sigma^n + \sum_j R_j^n}$$

where $R_i^n$ is neuron $i$'s driven input, $n$ is an exponent empirically estimated in the range of roughly 2–4, and $\sigma$ is a semi-saturation constant preventing division by zero at low input levels. Since the discovery, this neural circuit has been found in several other areas in the brain other than vision, including the motion processing in area MT, multisensory integration in parietal cortex, olfactory processing in the fly antennal lobe (Olsen & Wilson, 2008), attentional gain modulation (Reynolds & Heeger, 2009), and reward-value coding in parietal cortex (Louie, Grattan, & Glimcher, 2011).

This is directly relevant to the biological plausibility of categorical, competitive readouts of the kind used in this paper's discrete phoneme and character-level predictive coding pipelines. Heeger's own theoretical treatment of cortical function makes the connection explicit, noting that "max pooling (also called softmax) can be approximated by normalization" (Heeger, 2017). This means that the same divisive circuit already documented for contrast and motion processing is, in principle, capable of implementing the kind of competitive, winner-take-all-like selection among discrete alternatives that a softmax output layer performs in an artificial network.

One point worth noting here is that the biologically-supported nonlinearity in this circuit is a power law (i.e, response raised to exponent $n$), not the base $e$ exponential used in the standard machine-learning softmax function. The two forms are qualitatively similar since both produce graded, saturating competition among a pool of candidates, sharpening toward winner-take-all behavior as the exponent or gain increases, but they are not mathematically identical. The power-law form is the one with direct physiological support, while literal exponentiation appears in the literature mainly as a convenient link function in statistical (point-process GLM) fits to spike data rather than as an established biophysical mechanism.

### 5.4 Connection to Surprisal Theory, the N400 Effect and Garden Path Re-analysis

Over the years neuroscientists have discovered a number of patterns in the way the brain goes about processing language, and the three phenomena referenced in the header are the among the most well known among them. In the following we will investigate the relation between these and predictions from the IM-LEPP model.

#### Surprisal Theory

Hale (2001) and more influentially Levy (2008) proposed that the processing difficulty of a word during real-time comprehension is proportional to its surprisal, which was defined as the negative log probability of that word given its preceding context:

$$surprisal(w_i) = -\log p(w_i|w_{<i})$$

This equation says that predictable words (low surprisal) are read quickly while unexpected words (high surprisal) cause measurable slowdowns in reading time and eye-tracking measures. This has been extensively validated, including with surprisal estimates from modern LLMs. The IM-LEPP model provides a direct mathematical reason that explains the extra comprehension effort as explained next.

The IM-LEPP model uses a decomposition of a word $w$ into a sequence of characters $(ch_1, \dots, ch_k)$, so by applying the chain rule of probability, the surprisal definition becomes

$$surprisal(w) = -\log p(w|yy) = \sum_{j=1}^{k} [-\log P(c_j|c_1, \dots, c_{j-1}, yy)]$$

where the $yy$ is the latent representation of the word as predicted by the level 2 word level pipeline. Hence the generation of the word corresponding to $yy$ is decomposed into a serial generation of the characters for that word. Assuming that the ground truth for the $i^{th}$ character is given in 1-hot form by $t = (t_1, \dots, t_K)$, the previous section showed that the predictive coding pipeline for the $i^{th}$ character is driven by the minimization of the energy term $-\sum_k t_k \log y_k$ which reduces to the definition of surprisal $-\log y_{true\ character}$. If there is a big difference between the ground truth character $t$ and the predicted character (i.e. $y_{true\ character}$ is small), then the predictive coding energy pipeline will take longer to settle down for each character, and thus for the entire word, which maps naturally onto longer reading times, which is the surprisal effect.

Note that there is an additional error correction happening at level 2, where the level 1 generated ground truth latent representation of the word that actually arrived is compared with the predicted latent $yy$. This happens in continuous latent space, not over word identity, so it isn't measuring "how likely was this word" the way the summed level-1 terms do. It's closer to measuring "how well did the higher-level semantic/contextual expectation $yy$ fit what arrived," independent of how surprising the word's bare identity was, which brings us to the N400 effect.

**N400**

Kutas and Hillyard (1980) found a negative EEG deflection peaking around 400ms after a word was read, and this was larger for semantically unexpected words in the context. Their original example, "He spread the warm bread with socks," showed a large N400 for the anomalous final word. There's an active, unresolved debate in the ERP literature about what N400 actually indexes. Some accounts treat it as reflecting ease of lexical access/retrieval (a facilitation account), others as reflecting the difficulty of integrating a word's meaning into the evolving sentence/discourse representation (an integration account).

This maps remarkably well onto a distinction already built into IM-LEPP's two level structure for language: The summed level-1 cross-entropy terms are the natural computational candidate for the lexical-access side of the N400 debate (word-identity predictability, classical surprisal), while level 2's continuous prior-mismatch term is the natural candidate for the integration side (how well the word's meaning fits the evolving discourse representation). These are two terms that correspond to two different, independently-motivated quantities in the psycholinguistic literature, occurring at two different, architecturally-distinct levels. There's contemporary work pursuing exactly this kind of decomposition (a recent paper titled "Decomposition of surprisal: Unified computational model of ERP components in language processing," and Kuperberg, Brothers & Wlotko's work proposing distinct neural signatures for violated predictions at different levels of representation), so IM-LEPP's energy decomposition is relevant to a current research conversation rather than restating settled science.

Osterhout and Holcomb (1992) identified a distinct ERP component P600 corresponding to a positive deflection around 600ms which was specifically elicited by syntactic anomaly and distinct from N400's semantic profile. It's since been found to respond to a broader range of syntactic ambiguities and reanalysis demands, not just outright grammatical violations, which brings us to the next topic.

**Garden Path Re-Analysis**

A garden-path sentence is one that's temporarily structurally ambiguous in a way that leads readers to commit to an incorrect initial parse, requiring revision once disambiguating material arrives. The canonical example, from Bever (1970): "The horse raced past the barn fell". Readers initially parse "raced" as the main verb, then hit "fell" and must reanalyze "raced past the barn" as a reduced relative clause. Frazier and Rayner (1982) established the empirical signature via eye-tracking, which was increased reading times and regressive eye movements specifically at the disambiguating word 'raced' (is 'raced' the right verb for this sentence or is it 'fell').

Garden-path reanalysis can be explained using IM-LEPP as follows: On the first reading of the sentence, the diffusion based optimization causes the system latent state $xx_n$ at the output of the prediction module at level 2, to settle into a locally low-energy but incorrect interpretation after the word 'raced' is read, which propagates into a large error several words later when the word 'fell' is read. At this point the processing returns to the retained representation of 'raced' and this results in an escape from $xx_n$ to a new prediction $xx'_{n+j}$ that has a lower prediction error when the word 'fell' is read again ($j$ is the number of intervening words). The large prediction error encountered on the first reading of 'fell' changes the energy landscape and this facilitates the escape using the mechanism that explained in figure 8 in Varma.

The initial mis-interpretation of 'raced' connects to Ghio et al.'s (2024) finding that diffusion-based sampling can get trapped by an emergent competing local maximum along the annealing path, and this maps directly onto "the parser gets trapped in the wrong garden-path interpretation," with P600 as the plausible neural signature of the resulting escape-and-resettle process. "Garden-Path Model" is actually the name of a specific theory (Frazier's), proposing the parser commits serially to one analysis at a time. There's a rival, well-established account called the constraint-based/parallel tradition (MacDonald, Pearlmutter & Seidenberg, (1994)) which proposes that multiple candidate parses are entertained simultaneously with graded activation, more like a genuine probability distribution than a single serial commitment. IM-LEPP's own architecture, representing a distribution over candidate states via diffusion sampling before settling, is arguably a better structural match to the constraint-based account than to Frazier's original serial model, despite sharing the "garden path" name with the phenomenon both theories are trying to explain.

### 5.5 Comparison of Next Word Prediction Strategies in Transformers vs IM-LEPP

In lieu of doing experiments on the human brain, Barenholtz (2026) probed the internal dynamics of a transformer based LLM to understand how it generates language. He introduced a quantity called trajectory extrapolation error (TEE) which is defined as follows: At each word, fit a linear trajectory to the preceding 3 hidden states of a transformer (GPT-2/Pythia), extrapolate one step forward linearly, and measure how far the actual next hidden state lands from that extrapolated point.

His key finding is that TEE is nearly orthogonal to surprisal with $r = .044$ and independently predicts reading times. This result was replicated across GPT-2 sizes, on both garden-path sentences and thousands of naturalistic word positions (Natural Stories). The displacement control is the sharpest result: It says that raw magnitude of representational change (i.e. surprisal) and TEE predict reading time in opposite directions, so that a large change that continues the established direction is facilitative, while a small change that breaks direction is costly.

Critically, the GPT2 model itself doesn't use this trajectory structure in its own processing since direction preservation collapses to near-chance one step ahead at the intermediate layer. Thus each forward pass that results in the next word prediction is essentially a fresh computation, not a running extrapolation. Trajectory structure in GPT2 is a passive residual of training on human-produced text (which itself has local planning momentum), not an active computational strategy that the transformer exploits.

The core dissociation of word-level prediction error vs. representational reorientation cost maps onto the predictive-coding energy in the IM-LEPP model by using the energy equation

$$E_{PC} = \frac{1}{2}\epsilon_y^T \Pi_y \epsilon_y + \frac{1}{2}\epsilon_z^T \Pi_z \epsilon_z$$

Surprisal is the natural analogue of $\epsilon_y$ (bottom-up word-level error) while TEE looks like an empirical proxy for $\epsilon_z$ (top-down: how badly did the state land where its own momentum/prior predicted it should). Barenholtz found that these two are independent, additive, and non-interacting and is a nontrivial piece of evidence for keeping $\epsilon_y$ and $\epsilon_z$ as separate energy terms in the IM-LEPP model.

The paper explicitly finds that plain autoregressive transformers don't do what the design assumes, i.e. they don't carry forward a persistent state with real momentum that gets used prospectively. Instead they recompute fresh from context every step, and trajectory structure in their hidden states is an accidental byproduct of training data, not an active mechanism. Barenholtz himself flags this as the open question: Is trajectory-sensitivity in humans just a passive correlate of prediction, or is comprehension "fundamentally a dynamical process in which the evolving representational state carries local trajectory continuity that is actively maintained and exploited"? His data is consistent with either.

The IM-LEPP model is a specific, falsifiable bet on the second hypothesis. It is stronger than anything this paper establishes, but a natural sharpening of the exact question the paper leaves open. If anything, the paper's finding that an ordinary transformer's own dynamics don't do active momentum-tracking is a reason to think an ordinary transformer is the wrong underlying mechanism for what the human data show. A concrete follow-up test this suggests: replace the paper's linear extrapolation for the latent state (deliberately the crudest possible transition model) with an actual trained, ideally multimodal, diffusion based next- $z$ predictor, and check whether its residual explains reading times even better. That would be direct evidence favoring something like a diffusion-based transition module over a simple momentum model.

To summarize: The discussion of the Barenholtz paper lays bare the fundamental difference between how LLMs and the IM-LEPP model do language generation:

- LLMs sample from the distribution $p(w_n|w_{<n})$ at each step of the generation. Thus they are able to access all the data in their prefix, as well as everything that they have generated so far, which may run into hundreds of thousands of words for the latest transformer models.
- The IM-LEPP model generates its next state $xx_n$ by sampling using an energy function $E_W(x; zz_{n-1}, zzz_{n-1})$, it does not maintain all the previously data in its memory but instead summarizes its 'gist' in the form of the word latent state $zz_{n-1}$ and the hub state $zzz_{n-1}$ which gets translated into the next word.

It is quite likely that the brain operates like the IM-LEPP, since we don't keep the entirety of what we have read or heard while deciding what word to generate next. This also is a reflection of the difference between RNNs and transformers, since RNNs are a more primitive version of the IM-LEPP model (RNNs have a simpler linear model for prediction rather than diffusion based prediction). The full attention mechanism in transformers gives every past state a direct, undiluted, distance independent access, but it does come with some drawbacks: Liu et al. "Lost in the Middle' result shows that large context models don't use context evenly, they have U-shaped retrieval curve which is worse for information buried mid-context. In addition the self attention costs in transformers grows quadratically with context length.

Strictly speaking the brain uses another source of information while doing generation which we haven't put into the IM-LEPP model yet, and this memory. It is thought that some of the prior experiences that the brain has undergone gets stored as a sequence of latent states in the hippocampus, and gets recalled when the brain's language module is deciding on what word to generate next. The specific sequence of states that is recalled depends on the IM-LEPPs currently generated sequence of latent states $zz_n$ and is done using an associative memory mechanism such as the modern Hopfield network, this topic is further discussed in a following section. The prefix of already generated words $w_{<n}$ in transformers can be considered to be a type of short term memory with extremely large capacity.

## 6 Grounding, Data Efficiency, and Word Learning: Or How do Children Learn a Language Faster than LLMs

A striking empirical observation, increasingly emphasized in recent work comparing human language acquisition with transformer-based language models, is the enormous gap in data efficiency between the two: children acquire fluent, grammatical language from on the order of 100 million words of input by adolescence, while large language models typically require three to four orders of magnitude more data and still fall short of human performance on many measures (Warstadt, Choshen, Mueller, Wilcox, Zhuang, Linzen, et al., the BabyLM Challenge). This gap has motivated a dedicated research program (the BabyLM Challenge) aimed at training language models on human-scale, developmentally plausible corpora, and is closely related in spirit to the dissociation between raw next-word prediction accuracy and human-likeness discussed above in connection with Barenholtz (2026): models that achieve better next-word prediction do not necessarily become better models of human language processing, and by extension, the amount of data required to achieve human-level next-word prediction accuracy does not appear to be the relevant quantity for understanding how humans actually acquire language.

IM-LEPP suggests two distinct, complementary contributions to this data efficiency, both already implicit in the architecture developed above rather than requiring new machinery:

- **Grounding reduces the learning problem language must solve on its own.** As argued earlier in this paper, language in IM-LEPP does not construct its own latent representational space from word co-occurrence statistics alone; it reads from and writes to an already-existing, richly structured, amodal conceptual state at the central ATL hub, built up from whatever sensory and social experience is available prior to and alongside language acquisition. A transformer trained purely on text must recover an entire semantic space from distributional statistics alone, by contrast a language learner embedded in IM-LEPP's architecture is instead learning to map onto a semantic space substantially constructed by other means.

- **Word learning for an already-familiar concept requires only a one-sided update.** This can be made precise with a concrete example. Suppose a child has already developed a stable object-level model for apples through repeated non-linguistic visual experience, i.e., the vision-to-hub mapping for this concept is already well established. When a caregiver says "apple" while the child is attending to one, both the vision and language pipelines are active simultaneously, and both are trained against the same shared hub state at that moment. Vision learns that this percept is consistent with this region of hub-space, and language learns that this word is consistent with the same region. Critically, only the language-to-hub mapping is novel in this episode; the vision-to-hub mapping was already in place from prior experience. This is a direct mechanistic account of why word learning for a familiar referent is characteristically fast, one-trial or near-one-trial learning ( see "fast mapping," in Carey & Bartlett, 1978) since the system is attaching a second route into an already-populated region of representational space, not learning a new concept from scratch.

This account also makes a further prediction about what happens after the association is learned: hearing “apple” in the absence of any apple should push the shared hub state toward the same region, which then propagates back down to the vision pipeline as a top-down expectation. Depending on how much competing bottom-up sensory evidence is present, this produces one of two distinct effects, both independently well documented. With substantial competing visual input, the effect should manifest as priming, i.e., facilitated recognition of an actual apple if one subsequently appears, without generating a percept on its own, consistent with the extensive cross-modal semantic priming literature. With attention withdrawn from competing visual input, such as by closing eyes, or directing attention inward, the same top-down push can dominate sufficiently to drive genuine percept generation via $g_\psi$, consistent with evidence that visual mental imagery recruits much of the same visual cortex, including topographically organized early visual areas, as actual perception (see Kosslyn, Ganis, & Thompson, 2001). This is the same open-loop generative mechanism already invoked in Varma for planning, here triggered by a linguistic cue rather than a self-generated goal, with the choice between priming and full imagery governed by the same precision-weighting mechanism (Feldman & Friston, 2010) used throughout this paper for attention.

Grounding need not be specifically visual, and the evidence from congenital blindness is informative here. If language acquisition depended critically on visual grounding specifically, congenitally blind children would be expected to show substantial delays. The evidence does not support this in a simple way. Blind children correctly acquire and use vision-specific vocabulary, including verbs such as “look” and “see,” and color terms, from an early age in linguistically appropriate ways, and syntactic development proceeds on the same timeline as in sighted children (Landau & Gleitman, 1985)). At the same time, the picture is not a uniform null result since there is genuine variability across blind children in developmental timing, and there are specific narrow deficits. For instance, they have fewer words for objects that are visually but never tactilely accessible (such as flag or moon), and, in some studies, more articulation errors on speech sounds with visible lip and mouth configurations. This pattern is well accounted for by IM-LEPP’s architecture without modification: the hub is amodal and integrates whichever spokes are available, so grounding is not architecturally tied to vision specifically, and broad semantic and syntactic development can proceed normally by drawing on other available modalities (audition, touch, social interaction). Deficits appear only in the narrow cases where no other modality can substitute for genuinely vision-specific content. Notably, recent work further suggests that blind children’s language input is not systematically different from that of sighted children (Campbell, Righter, Lukin, & Bergelson, 2025), consistent with compensation occurring naturally through however caregivers route grounding-relevant information through the channels actually available, rather than through any special-purpose mechanism specific to blindness.

## 7 Inclusion of Memory

There are several places in the paper where we pointed out the need for a memory sub-system for the IM-LEPP model which differs from the full attention mechanism used in transformers. The need for a memory sub-system is also motivated by Lewis and Vasishth's (2005) activation and cue based retrieval model that explains long distance dependency resolution in the brain not via continuous attention over everything, but via content addressable retrieval from a declarative memory, cued by partial feature matches and modulated by decay and similarity based interference. Memory based designs have also been proposed to make transformers more efficient, for example see Borgeaud et al. (2022).

There are two kinds of memory sub-systems that have been alluded to in this paper:

- **Object Model semantic memory:** The level 2 vision hub interfaces with predictive processing modules that are instantiated on a per-object basis, depending on the objects that are present in the field of vision, and which are consciously being tracked. Clearly it is not efficient to build up the predictive model for each object (involving the parameters for the energy models $q_\phi$, $g_\psi$ and $E_W$) from scratch every time the object appears, and hence points to a system in which the brain stores away object models for later use, perhaps in the brain's cortical columns. These models are recalled and connected to the level 2 vision hub when the time arises, and as a result their parameters evolve while they are active, and once they are no longer needed, they are once again stored away with the modified parameters. Hence, as has been pointed out by others, recalling a memory is an active process, in which the contents of the memory changes as a result of the recall. This memory is probably organized in an hierarchical fashion, for example there could be model for the generic category of dogs, while under it there could be specific models for various breeds of gods, and at a lower level, a model for my pet dog.
- **Episodic Memory:** Each stored element in this system consists of a sequence of latent states $zzz_1, \ldots, zzz_N$ sampled from the central ATL hub. Sampling states from the ATL ensures that these states are multimodal, so that they can be turned into a sequence of images or they can be turned into sound, language or other modalities. Tulving in a classic work from 1972 proposed that episodic memory is inherently events extended in time. There is a fair amount of evidence that the seat for episodic memory in the brain lies in the hippocampus. There is direct physiological evidence that hippocampal place cells replay compressed, ordered sequences (see Wilson and MacNoughton (1994)).

Vargha-Khadem, Gadian, Watkins, Connelly, Van Paesschen & Mishkin's (1997) classic study of developmental amnesia in which they studied children with early, selective bilateral hippocampal damage, found severely impaired episodic memory alongside broadly age-appropriate semantic/factual knowledge acquisition throughout development. This shows a genuine double dissociation, supporting the need for both object level semantic storage as well as as episodic memory.

There are several questions that needs to be answered in the design of a memory sub-system:

- What is the nature of the memory to be used? In all likelihood an associative memory of the modern Hopfield network type is an appropriate choice.
- What is the storage criteria? Obviously not all episodes that we come across with need be stored, nor do we need to store models for objects that we almost will never comes across a second time, such as strangers on the street. Some possible storage criteria ideas include: Store episodes whose state prediction $x_n$ differs a lot from the state $z_n$ that results from new ground truth data. This corresponds to a state of surprise in human terms, and should result in the storage of $x_n$ and the subsequent states in to episodic memory. This criteria applies equally well to vision or to language data and in the case of language it results in the storage of thoughts that are of significance. The length of the sequence to be stored is governed by when the next surprise state occurs, at which point it is terminated. This is reminiscent of the design used for splitting up a sequence of characters into words in the language model, in this case we are splitting up a sequence of words into thoughts, with the caveat that not all thoughts are stored, only those whose surprise value exceeds some threshold. This matches a real biological mechanism that was proposed by Lisman and Grace (2005). They proposed a hippocampal VTA loop for detection of newly arrived information not already stored, by triggering a novelty signal and dopamine release, which is like prediction error gated writing in our model. Another storage criteria can be developed based on the valence signal coming from the amygdala. It is possible that the amygdala gets activated as a result of the surprise value alluded to earlier, and it generates the valence signal for the ATL hub that leads to the storage of the episode. In this sense the amygdala's signal will not be a separate mechanism but a way in the surprise mechanism is implemented by the system at the ATL hub.
- Since it is an associative memory, what are query and key values to be used? What are the rules to be used for memory decay and over writing?
- How many episodes or objects can be active simultaneously at any one time? Empirical data from human tests shows that the capacity is about 3-4 chunks (see Cowan (2001))
- What is the mechanism for feeding episodic memory states into the system? The most natural point of entry is the central ATL hub. The ATL itself is connected to the hippocampus through a rich set of connections. A study using intracranial recordings during a word-list learning task found that hippocampal sharp-wave ripples coordinate with cortical ripples specifically in the ATL-centered semantic network during both encoding and recall. Since sharp-wave ripples are the neural signature of hippocampal replay (see Wilson & McNaughton, 1994), this shows that exact replay machinery interacting, in a temporally precise way, with the ATL hub specifically during real memory formation and retrieval. Hippocampal replay is time compressed at a speed that is 10-20x faster than originally recorded. Thus an entire memory episode can be replayed in the time between successive state updates at the word level.

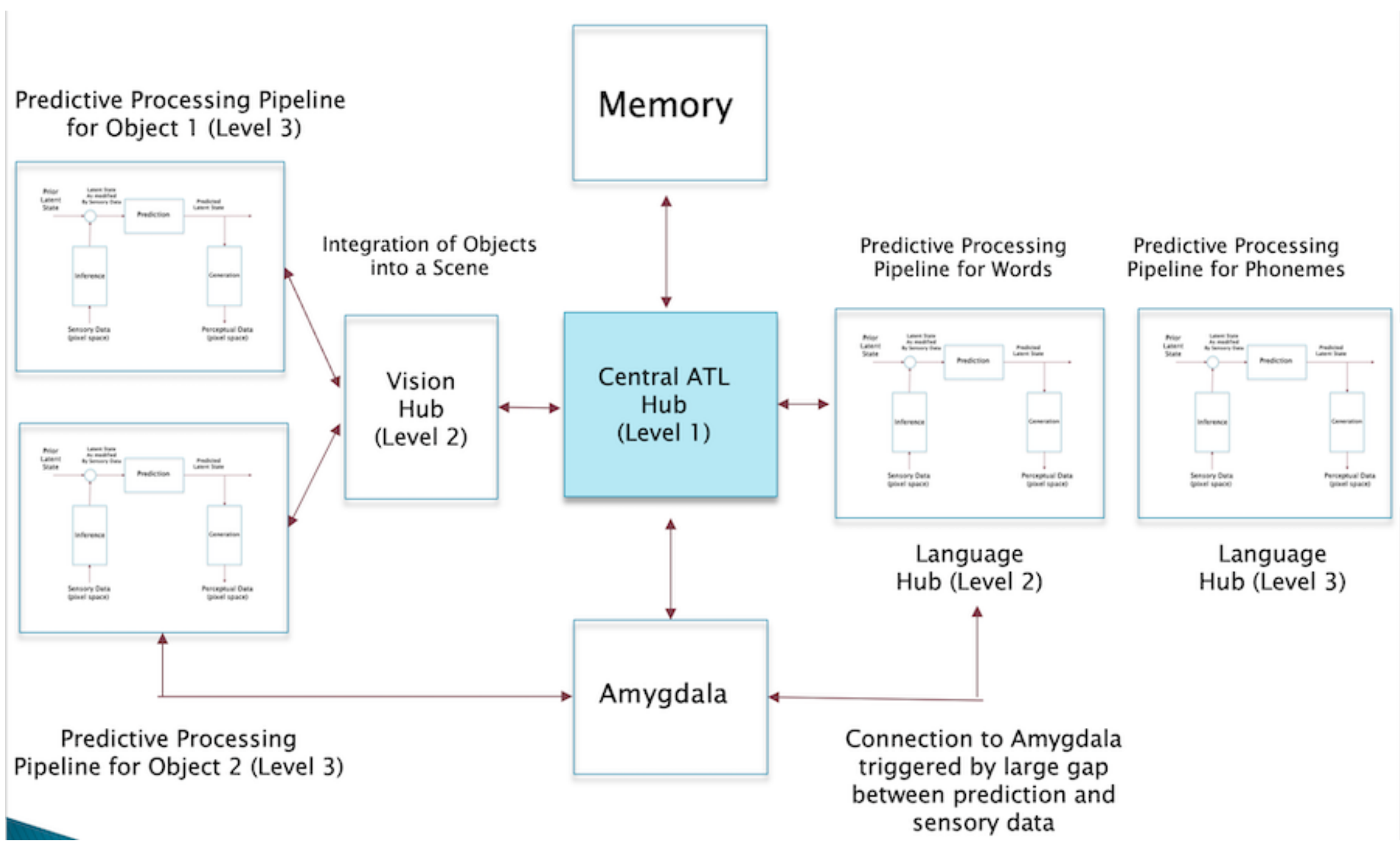

Figure 14: Incorporation of Episodic Memory into the IM-LEPP system

The above figure shows a proposed design for incorporation of episodic memory into the IM-LEPP model. It shows connections between the predictive processing modules and the amygdala, for both vision and language. When the prediction module in any of these pipelines notices a large gap between the prediction $x_n$ and the sensory data mediated next state $z_n$, then it sends a signal to the amygdala. Subsequently the amygdala communicates with the ATL hub, which initiates the process of storing the hub state $xxx_n$ into the memory. The hub states continue to be stored until the reception of the next signal from the amygdala.

We are working on the details of this model, and will be documented in an upcoming paper.

## 8 Existing Work

A number of existing theoretical frameworks propose that the brain operates via hierarchical predictive processing and energy or free-energy minimization. This section situates IM-LEPP relative to the most directly relevant of these, organized by research program rather than chronologically, since several of these lines of work span decades and continue to develop in parallel.

**Foundational predictive coding.** The inference-prediction-generation cycle used throughout this paper builds directly on Rao and Ballard's (1999) original formulation of predictive coding in visual cortex, which proposed that top-down connections carry predictions of lower-level activity while bottom-up connections carry only the residual prediction error. IM-LEPP extends this single-level, single-modality, purely continuous formulation in several directions developed throughout this paper: extension to a diffusion based temporal prediction mechanism (originally done in the earlier LEPP model), a genuinely hierarchical hub-and-spoke architecture (rather than a fixed chain of levels), an extension to discrete/categorical sensory data and multimodal integration via a central amodal hub.

**Friston's free-energy principle and active inference.** Friston's (2010) free-energy principle generalizes predictive coding into a much broader theoretical program, proposing that essentially all adaptive biological self-organization, not just perceptual inference, but action selection, learning, and homeostasis, can be understood as approximate minimization of variational free energy, with perception and action both serving to reduce the same quantity (Friston and Kiebel, 2009; Friston, Parr, & Pezzulo, 2022). This is a substantially more ambitious unification than IM-LEPP attempts since active inference treats action as inference over policies that minimize expected free energy, giving a principled account of behavior that IM-LEPP, which focuses specifically on perceptual and linguistic representation, does not currently address. Friston's own attempt to extend predictive coding into the temporal domain in the form of hierarchical dynamic models using generalized coordinates of motion, in which each level represents not just a state but its temporal derivatives (Kiebel, Daunizeau, & Friston, 2008), is one of the specific proposals Heeger (2017) critiques as inconsistent with the sustained-activity evidence discussed earlier in this paper. IM-LEPP's diffusion-based temporal prediction module is a different, later, and more explicitly probabilistic solution to the same problem Friston's generalized coordinates were designed to solve, however it represents genuine uncertainty and multimodality in the predicted next state, rather than a deterministic trajectory in an extended state space.

**Heeger's Theory of Cortical Function.** Heeger's (2017) framework shares IM-LEPP's core mathematical structure of neural dynamics as gradient descent on an energy function combining a feedforward and a prior term, but explicitly rejects the Rao-Ballard direction of information flow, on the grounds that predictable stimuli produce sustained rather than vanishing activity. Heeger's temporal prediction mechanism (implemented using banks of quadrature-phase oscillator pairs) is linear and comparatively lightweight relative to IM-LEPP's diffusion-based module, and his account of perceptual multistability (persistent, non-vanishing noise under a fixed landscape) is a mechanism this paper has explicitly incorporated for Necker-cube-style switching, while retaining landscape-reshaping via new evidence for cases such as garden-path reanalysis. Subsequent work in this program (ORGaNICs Heeger & Mackey, 2019), which extended the framework to working memory and motor control, and an explicit candidate circuit for divisive normalization (Heeger & Zemlianova, 2020)), remains the closest existing body of work to IM-LEPP in its underlying mathematics. The point of disagreement over sustained versus decaying activity is a genuine, unresolved empirical question this paper has not settled, only clarified.

**Temporal predictive coding (tPC).** Millidge, Tang, Osanlouy, Harper, and Bogacz (2024) propose a state-space predictive coding model that, like IM-LEPP, processes input incrementally and maintains a persistent latent state updated via local prediction-error minimization, and this is architecturally the closest existing model to IM-LEPP's core per-object and per-word inference-prediction cycle. Thus tPC is a counterpart to the LEPP model from Varma. Ruini et al. (2025) apply this framework directly to online sentence processing, making it directly comparable to the language model developed in this paper. The key architectural difference is the transition function: tPC uses a linear-Gaussian state transition, which — as this paper has argued at length in connection with the averaging/blurring problem in early image-generation models — cannot represent genuinely multimodal beliefs about the next state. IM-LEPP's diffusion-based prediction module is proposed specifically to address this limitation, at the cost of the considerably heavier computational machinery.

**Joint-Embedding Predictive Architecture (JEPA).** LeCun's (2022) "A Path Towards Autonomous Machine Intelligence" proposes predicting in latent representation space rather than reconstructing raw sensory input, framed via an energy function scoring the compatibility between a context representation and a predicted target representation and is the closest existing premise to IM-LEPP's own commitment to operating on learned energy landscapes rather than reconstructed sensory detail. The published instantiations, I-JEPA (Assran et al., 2023) for static images and V-JEPA (Bardes et al., 2024) for video, predict masked or future latent patches from context, without pixel-level reconstruction or hand-crafted augmentations. Two differences are worth being explicit about: First, despite the shared energy-based framing, published JEPA predictors are comparatively lightweight, near-deterministic networks trained with a direct latent-space regression loss, not genuine stochastic sampling over a learned energy landscape — the same limitation already discussed above for tPC's linear-Gaussian transition, and the reason IM-LEPP's diffusion-based $E_W$ (capable of representing genuinely multimodal beliefs about the next state, including the perceptual bistability discussed for the Necker cube) is a heavier but more expressive mechanism than either. Second, JEPA-family methods generally require additional architectural safeguards, typically an asymmetric or momentum-updated target encoder, to prevent the representational collapse that a purely latent-space predictive objective is prone to. IM-LEPP's inference network is anchored by an explicit reconstruction term, i.e., a generated percept compared against real sensory data, that JEPA's purely latent-space objective specifically forgoes. This rules out the degenerate constant-embedding solution JEPA's momentum-encoder machinery is designed to prevent, though it does not make IM-LEPP immune to the different, better-studied posterior-collapse failure mode familiar from the VAE literature. LeCun's broader H-JEPA blueprint does anticipate eventual multimodal integration and a hierarchical, multi-timescale architecture in the spirit of IM-LEPP's own hub-and-spoke design, but the published, concrete JEPA systems to date remain single-modality, and none carries IM-LEPP's specific commitment to a biologically-grounded ATL hub-and-spoke architecture or its contact points with human psycholinguistic data.

**Hierarchical Temporal Memory and the Thousand Brains Theory.** Hawkins and colleagues propose a substantially different architecture built around thousands of largely identical, quasi-independent cortical columns, each learning a complete sensorimotor model via grid-cell-like reference frames, with perception emerging from voting/consensus across columns rather than convergent hierarchical integration (Hawkins & Ahmad, 2016; Hawkins, Lewis, Klukas, Purdy, & Ahmad, 2019). This shares IM-LEPP's premise that cortex is fundamentally predictive, but differs in nearly every other respect, since there is no energy function or Bayesian formalism, a flat rather than hierarchical integration structure, and a different representational currency (modular grid-cell-like codes rather than continuous latent vectors). It is a useful alternative to keep in view specifically for the multi-object, multi-region integration questions this paper addresses, since it represents a genuinely different answer to the same architectural question, i.e., many parallel, redundant, voting modules rather than a small number of specialized converging hubs.

**Predictive coding as an alternative to backpropagation.** A separate strand of the predictive coding literature, exemplified by Whittington and Bogacz (2017), is primarily concerned with predictive coding as a biologically plausible local learning rule approximating backpropagation, rather than as a model of perceptual or cognitive dynamics per se. This paper draws on the same local, gradient-based update rules, but for a different purpose — as a proposed account of real-time inference and prediction, not primarily as a training algorithm — and this distinction is worth keeping explicit, since the two research threads are often cited together despite addressing different questions.

**The broader predictive processing program.** Beyond the specific computational models discussed above, predictive processing has been developed as a general framework for the mind in philosophy of cognitive science, notably by Clark (2013) and Hohwy (2013). IM-LEPP can be read as a specific, mechanistically detailed proposal within this broader program, distinguished by its commitment to energy-based/diffusion dynamics, its explicit multimodal hub-and-spoke architecture, and its attempt to derive contact with quantitative psycholinguistic data (surprisal, N400, garden-path reanalysis) from the model's own mathematical structure rather than from analogy alone.

Taken together, IM-LEPP's specific position relative to this literature can be stated as follows: It is closest in spirit and mathematics to Heeger's framework and to tPC, while retaining Rao-Ballard-style vertical error propagation that Heeger's account explicitly rejects, it addresses a temporal-prediction problem that Friston's generalized coordinates and tPC's linear transition both address with comparatively weaker transition models and finally it proposes a converging, sharing JEPA's premise of predicting in latent rather than pixel space, while offering a more expressive (diffusion-based, genuinely stochastic) transition mechanism than either tPC or published JEPA predictors provide and finally a hierarchical, multi-hub integration architecture as a specific alternative to Hawkins' flat, voting-based one.

## 9 Conclusions

This paper has extended the LEPP framework for perception into IM-LEPP, as a hierarchical, multimodal, energy-based model of cognition integrating vision and language through a hub-and-spoke architecture grounded in Lambon Ralph et al.'s controlled semantic cognition framework. Building on the shared claim that generative neural networks may be understood as effective theories of cognitive dynamics operating at the level of learned energy landscapes, we have proposed a hierarchical hub-and-spoke integration architecture for multiple sensory modalities; a multi-level vision model with mechanisms for attentional suppression, saliency-gated pipeline instantiation, and open-loop prediction; a predictive processing model for language spanning phoneme and character-based input, with an extension of predictive coding to discrete/categorical data and a biologically plausible candidate implementation of the categorical readout via divisive normalization; and an outline of a memory subsystem distinguishing semantic and episodic storage. We have further shown that the model's own mathematical structure recovers or motivates a range of independently established findings in psycholinguistics and neurolinguistics, namely surprisal theory, the N400 and P600 components, garden-path reanalysis, and a specific point of contrast with transformer-based language models regarding trajectory-sensitivity in next-word prediction.

**Future Extensions**

**A complete treatment of the memory subsystem.** As noted above, the memory design presented here is intentionally incomplete: the precise specification of query and key representations, decay and overwriting rules, and the full mathematical treatment of the underlying modern Hopfield network formalism are left to a dedicated future paper, which should also address the relationship between the object-model (semantic) and episodic memory systems more precisely than has been possible here.

**The remaining spokes of Figure 1.** This paper has developed models for vision, speech, and valence, leaving two of Lambon Ralph et al.'s six proposed spokes, namely praxis and function, as natural directions for extension. Praxis concerns motoric and action-related knowledge: the specific manner in which an object is manipulated (for instance, that a steering wheel affords rotation), associated in the empirical literature with parietal regions such as the intraparietal sulcus and inferior parietal lobule.

A praxis pipeline within IM-LEPP would most naturally be modeled as a predictive processing pipeline over motor sequences, predicting and generating intended actions rather than percepts and could plausibly reuse the state feedback control machinery already introduced in this paper for speech production (Hickok, Houde, & Rong, 2011), generalized from the specific case of articulation to motor action more broadly, since both are instances of the same general problem of predicting the sensory consequences of an intended movement and correcting against feedback.

Function, by contrast, concerns more abstract knowledge of what an object is for, which the neuropsychological literature suggests is dissociable from praxis (patients can show selective deficits in one while sparing the other) but closely coupled to it. Rather than introducing an entirely separate pipeline, function knowledge may be more naturally modeled as an additional property learned within the existing object-model semantic memory described in this paper and thus part of what a stored object model comes to predict about an object's typical interactions and uses, rather than a wholly independent spoke. A further, related extension worth noting is that Lambon Ralph et al.'s framework distinguishes general environmental sound (for instance, recognizing a dog by its bark) from speech specifically; this paper has developed only the speech/phoneme pathway, leaving a hub-and-spoke treatment of non-linguistic auditory object recognition, analogous to the vision system's object-tracking architecture, as a further natural extension.

**A working implementation.** The most ambitious extension, and a prerequisite for much of the experimental program outlined below, is an actual trained instantiation of some portion of this architecture, even a reduced-scale version, such as a multi-object visual tracking and scene-integration task, or the language model applied to a moderately sized text corpus, that is sufficient to test whether the qualitative behaviors this paper predicts (attention-gated inattentional blindness, saliency-gated instantiation delays, garden-path reanalysis dynamics) actually emerge from trained dynamics rather than being read into the architecture by construction.

### Experimental Predictions

Several of the mechanisms proposed in this paper make specific, checkable predictions, distinct in character from the psycholinguistic phenomena already discussed, which this paper has shown to be consistent with rather than independently confirmed by the model.

The most immediately tractable is a direct extension of the comparison with Barenholtz (2026) discussed above: training an actual nonlinear, ideally diffusion-based, next-state predictor on transformer hidden states and testing whether its residual explains human reading times better than Barenholtz's linear extrapolation baseline. This test uses only existing models and existing reading-time datasets, and would provide direct evidence for or against the model's central claim that genuine multimodal prediction, rather than simple momentum, underlies representational reorientation costs in comprehension.

A second test targets the account of the N400 proposed above: fitting the summed level-1 cross-entropy term and the level-2 prior-mismatch term separately (and jointly) as predictors of measured N400 amplitude in existing ERP datasets, to determine whether the lexical-access and integration accounts debated in that literature correspond to these two architecturally distinct quantities, as the model predicts, or whether one alone suffices.

A third test concerns the dwell-time statistics proposed for perceptual multistability: the model predicts non-memoryless, gamma-distributed dominance durations for genuinely ambiguous stimuli, with systematically asymmetric dwell times for stimuli with an ecological prior (such as the Necker cube's viewing-from-above bias), arising specifically from unequal depth between the two energy minima rather than from asymmetric state persistence. This is directly testable against existing and new bistable-perception datasets by comparing the predicted dwell-time distribution shape against measured data.

A fourth test concerns the memory write-gating criterion: recognition-memory experiments that orthogonally manipulate prediction-error magnitude and emotional valence could determine whether episodic memory strength follows the specific combination rule this paper's architecture implies, rather than merely confirming that both factors independently matter, which is already well established.

Finally, and most broadly, an actual trained implementation of the kind described above would allow direct comparison of the model's emergent behavior against the qualitative phenomena discussed throughout this paper, namely inattentional and change blindness, garden-path reanalysis dynamics, and the multi-rate integration of asynchronous sensory streams. This will provide a considerably stronger test than consistency with independently established findings alone.

### Closing Remarks

The central claim of this paper, and of the LEPP framework it extends, is that generative neural networks operating via energy minimization may serve as effective theories of cognitive dynamics, in the same sense that statistical mechanics serves as an effective, non-mechanistic description of thermodynamic phenomena. This paper has aimed to show that such a framework, extended hierarchically across levels and modalities, is compatible with a substantial and heterogeneous body of existing evidence, which extends from the neuroanatomy of the semantic hub-and-spoke system to the fine-grained statistics of eye movements during garden-path reanalysis. Compatibility with existing evidence is a necessary condition for a theory of this kind, but it is not sufficient, and the experimental program outlined above is offered as a concrete path toward the stronger forms of evidence, namely direct quantitative tests and eventually a trained working implementation, that would be needed to move this framework from a coherent hypothesis to a validated model of cognition.